\documentclass[prd,twocolumn,superscriptaddress,floatfix,amsmath,amssymb]{revtex4-1}
\usepackage[utf8]{inputenc}
\usepackage{graphicx}
\usepackage{hyperref}
\usepackage{color}
\usepackage{bm}
\usepackage[normalem]{ulem}

\usepackage{pgfplots} 
\usepackage{tikz}
\usetikzlibrary{arrows,graphs,positioning}
\usetikzlibrary{decorations.pathreplacing}

\newcommand{\integral}[3]{\int_{#2}^{#3} \!\! \mathrm{d} #1 \,}

\newcommand{\pv}{\text{p.v.}} 

\newcommand{\ee}[1]{\mathrm{e}^{#1}} 

\newcommand{\ket}[1]{\left| {#1} \right\rangle}

\newcommand{\ketbra}[2]{\left| {#1} \right\rangle\!\left\langle {#2} \right|}

\newcommand{\intH}{H_{I}}

\newcommand{\ii}{\mathrm{i}}
\newcommand{\tr}{\operatorname{Tr}}

\newcommand{\normord}[1]{\mathop{:}\nolimits\!#1\!\mathop{:}\nolimits}

\newcommand{\exptval}[1]{\left\langle {#1} \right\rangle}
\newcommand{\comm}[2]{\left[{#1},{#2}\right]}

\begin{document}

\title{Information travels in massless fields in 1+1 dimensions where energy cannot}

\author{Robert H. Jonsson}
\affiliation{Department of Applied Mathematics, University of Waterloo, Waterloo, Ontario, N2L 3G1, Canada}

\begin{abstract}
It has been demonstrated that, in (1+1) spacetime dimensions, massless fields can be used for information transmission even between parties that are time-like rather than light-like separated - and even without the receiver obtaining any energy from the sender.
Here, it is shown that this phenomenon is not limited to a particular model of signaling device, but is based on general properties of the quantum field: energy propagates strictly only on the lightcone, whereas perturbations to the field amplitude, which can carry information, permeate also the inside of the future lightcone of the sender.
It is also shown that this timelike and energyless signaling in massless fields occurs in Dirichlet cavities,  which shows that  the phenomenon is not confined to setups with a zero mode.  
Moreover, it is shown that the phenomenon extends beyond perturbation theory, namely by deriving the phenomenon non-perturbatively for the case where the sender and receiver systems are modelled as harmonic oscillators. 
To illustrate the propagation of energy in massless fields, the energy density is calculated that results from temporarily coupling an Unruh-DeWitt detector at rest to the vacuum of the field.
\end{abstract}

\maketitle

\section{Introduction}

The study of the propagation and processing of information in quantum fields today plays an important role in several areas of physics, reaching from quantum gravity to experimental quantum information processing.

For example,  new insights into the interplay of quantum theory and gravity are emerging from the black hole information paradox \cite{mathur_information_2009}, and its proposed resolutions, such as the recent firewall conjecture \cite{braunstein_better_2013,almheiri_black_2013}.


Other research  has explored how the impact of spacetime curvature and relativistic acceleration on quantum fields can be quantified in terms of information-theoretic notions, such as channel capacities,  through the study of communication via quantum fields  \cite{bradler_quantum_2012,cliche_relativistic_2010,downes_quantum_2013}. 

At the same time, recent progress in experimental quantum information processing, e.g., in cavity QED and superconducting circuits, makes it possible to experimentally study  impacts of relativistic effects on  quantum fields \cite{wilson_observation_2011,friis_relativistic_2013,lindkvist_twin_2014} that were not previously observable.

Also, a feature of quantum fields that is now attracting particular interest, is their quantum vacuum fluctuations. 
For example, the seminal work \cite{reznik_violating_2005} showed that entanglement in the vacuum correlations of a quantum field can be swapped to devices coupling to the field, even if the devices are spacelike separated.
Following this result, it was shown that different spacetime geometries and relativistic observer trajectories can be distinguished by their entangling characteristics \cite{martin-martinez_entanglement_2014,steeg_entangling_2009,cliche_vacuum_2011,salton_acceleration-assisted_2015}.
Furthermore, the entangling of quantum systems coupling to the vacuum of a quantum field has been investigated as  a source of entanglement for use in quantum information processing \cite{brown_thermal_2013,martin-martinez_sustainable_2013}, and an experiment for the extraction of entanglement from the field has been put forward \cite{sabin_extracting_2012}.

Another surprising effect that uses the quantum correlations of the field is  quantum energy teleportation \cite{hotta_quantum_2014,hotta_quantum_2008,hotta_controlled_2010,verdon-akzam_asymptotically_2016}: In this protocol, two parties can teleport energy from one to the other, using only local operations and classical communication.

In this general context, the present article follows up on the recent observation that the flow of information in a massless field can be decoupled from the flow of energy to a large extent \cite{jonsson_information_2015}, or even entirely in certain circumstances. 
This phenomenon is a consequence of the violation of the strong Huygens' principle in general curved, and in  (1+1)- and in ($2n$+1)-dimensional Minkowski space \cite{mclenaghan_validity_1974,czapor_hadamards_2007,poisson_motion_2011}. When the strong Huygens' principle is violated, signals do not propagate strictly at the speed of light in a massless field, but part of the signal travels slower and reaches into the future lightcone of the sender. Crucially, this timelike part of the signal, in spite of being able to carry information, then tends to carry relatively little  energy (and in some circumstances even none). This is because the field amplitude, which is what is carrying the signal, does not contribute to the energy density of a massless field directly. Instead, only (the relatively small) derivatives of the field inside the forward light cone contribute to the field's energy density there.
This phenomenon occurs, for example, in cosmological models, and could have interesting implications for obtaining information about the early universe \cite{blasco_violation_2015,blasco_timelike_2016}

The effect is most pronounced in (1+1)-dimensional Minkowski space: Here, in fact, information can be transmitted between timelike separated parties via a massless quantum field without any energy propagating from the sender to the receiver. Both parties still have to expend energy locally on their interaction with the field to send, and to receive a message. The timelike signals are energyless though, in the sense that none of the energy the sender injects into the field is transferred to the receiver.

In \cite{jonsson_information_2015}, the effect was studied modeling the signaling devices as Unruh-DeWitt detectors. These are a theoretical model for particle detectors  detectors \cite{unruh_notes_1976,hawking_quantum_1979} that are an important theoretical tool for the study of quantum fields in curved spacetime, and are widely used in the field of relativistic quantum information, including many of the works cited above. The model is a good approximation to the interaction between realistic atoms and the electromagnetic field in microwave, and optical cavities \cite{martin-martinez_wavepacket_2013}. It also underlies the Jaynes-Cummings model, which is restricted to non-relativistic scenarios \cite{benincasa_quantum_2014,jonsson_quantum_2014}, and can be obtained from the Unruh-DeWitt model through  single-mode, and rotating-wave approximations.

In the present paper we show that timelike, energyless signals are not limited to a particular model of signaling device. Instead, as we discuss, they result from the way in which signals and energy propagate in the field, and we show that they appear in a wide range of scenarios.

To this end, in Section \ref{sec:comm}, we begin with a brief review of signaling in quantum fields, and the central role which the field commutator plays in signaling. Section \ref{sec:propagation} discusses how the energy density of a scalar massless field can be expressed through operators acting only on the boundary of its past lightcone, whereas the field amplitude depends on operators located inside the past lightcone as well. This provides the fundamental explanation
 for the phenomenon of timelike, energyless signals.

In Section \ref{sec:energyfromdet} we  calculate the  energy density injected into the vacuum state of the field by an Unruh-DeWitt detector which is coupling to the field for a finite time. We observe that the energy propagates strictly at the speed of light, in accordance with the findings from the preceding section.

A well-known problem of massless fields in (1+1)-dimensional spacetimes is the infrared divergence or ambiguity of the two-point function.
To show that timelike signaling is not related to this issue, in Section \ref{sec:cavity}, we study the commutator of a massless field in a cavity with Dirichlet boundary conditions, which is finite in the infrared and has no zero mode.

In prior work \cite{jonsson_information_2015,blasco_violation_2015,blasco_timelike_2016}, timelike signals where only discussed within  perturbative analysis. Therefore, in  Section \ref{sec:honew}, we show that they also appear in a non-perturbative treatment of signaling between harmonic oscillators coupling to the field inside a cavity. This model is, on one hand, closer to realistic experimental scenarios. On the other hand, this observation also opens up the interesting prospect of applying continuous variable quantum information methods to the study of timelike, energyless signals.

These findings raise questions about  the  relation between energy and information transport in quantum fields, such as how much quantum information energyless signals can maximally carry. They also strongly motivate the investigation of an experimental demonstration of the effect, for example, in superconducting circuits.

We use natural units, $c=\hbar=k_B=1$,  throughout the paper.

\section{Signaling in quantum fields}\label{sec:comm}
Our ability to send information across spacetime via a quantum field is governed by the commutator of the field: 
If and only if the commutator of the field is non-vanishing communication is possible.

If the commutator between two points in spacetime vanishes, 
then a sender interacting with the field locally at one point has no influence which could be used to send a signal to a receiver interacting with the field locally at the other point \cite{cliche_relativistic_2010,eberhard_quantum_1989}. If, however, the commutator is non-vanishing, then the two parties can use simple systems such as Unruh-DeWitt detectors \cite{hawking_quantum_1979,unruh_notes_1976} as signaling devices to send information from one point to the other 
\cite{cliche_relativistic_2010,jonsson_quantum_2014,jonsson_information_2015}.

This also shows that communication via a quantum field is possible between the same points as via the corresponding classical field, because the commutator
\begin{align}
\comm{\phi(x)}{\phi(y)}=\ii G(x,y)=\ii\left(G_{adv}(x,y)-G_{ret}(x,y)\right)
\end{align}
is given by the classical Green's function \cite{fulling_aspects_1989}.
 
For a massless Klein-Gordon field in (3+1)-dimensional Minkowski space, this means that signals propagate strictly at the speed of light, because the commutator is supported only on the null boundary of the lightcone.
\begin{align}
\comm{\phi(x)}{\phi(y)}&= \frac{\ii}{4\pi\left|\vec{x}-\vec{y}\right|} \left( \delta\left(x^0-y^0+\left|\vec{x}-\vec{y}\right|\right) \right. \nonumber\\
&\qquad\qquad\qquad\left. - \delta\left(x^0-y^0-\left|\vec{x}-\vec{y}\right|\right) \right)
\end{align}
This property, which is also called the strong Huygens' principle, however, is not a general property of massless fields. It is a very special property of flat, (3+1)-dimensional spacetime. On general, curved spacetimes, but also in (2$n$+1)-dimensional Minkowski space Huygens' principle is violated, and the commutator is non-vanishing even at timelike separations \cite{mclenaghan_validity_1974,czapor_hadamards_2007,poisson_motion_2011}.
When this is the case, signals can travel slower than light, contrary to the intuition based on (3+1)-dimensional Minkowski space.
(For a detailed calculation of the commutator in Minkowski space of different dimensions see, e.g., \cite{martin-martinez_causality_2015}.)

A particularly interesting case is (1+1)-dimensional Minkowski space. Here the commutator is  constant inside the lightcone, where it takes the value $\comm{\phi(x)}{\phi(y)}=\ii/2$ whenever $y$ is in the future lightcone of $x$:
\begin{align}
\comm{\phi(x)}{\phi(y)}=\frac{\ii }2 \Theta\left((x-y)^\mu (x-y)_\mu\right) \mathrm{sgn}\left(y^0-x^0\right) 
\end{align}
where $\Theta(t)$ denotes the Heaviside step function, and $x^\mu y_\mu= x^0 y^0-x^1 y^1$ is the Minkowski scalar product.
This makes it possible to send signals via the field to arbitrarily distant, timelike separated receivers.

A surprising property of these timelike signals in (1+1)-dimensional Minkowski space is that while they carry information information, they carry no energy from the sender to the receiver.
This was shown for a communication scenario with Unruh-DeWitt detectors as signaling devices in \cite{jonsson_information_2015}. It was found, within a perturbative analysis,  that any change in the energy of the receiver's detector due to the signal is instead accounted for by the work required of the receiver to couple and decouple his detector and the field.

The phenomenon of timelike, energyless signals is however independent of the particular signaling device. Instead it relies on the different propagation behavior of energy density, and the field amplitude as we discuss in the following.

\section{Signal and energy propagation in 1+1D Minkowski space}\label{sec:propagation}
In this section we review the propagation of energy, and field amplitude in a massless Klein-Gordon field in (1+1)-dimensional Minkowski space. 
Here, the energy density of the field is just the sum of the energy density of the so called left- and right-moving sectors of the field, which propagate strictly at the speed of light. We will see this below by expressing the energy density operator at a given point in spacetime, in terms of field operators acting on the field at earlier times.
This explains, why energy injected into the field by any signaling device always propagates away at the speed of light. So if a receiver wants to collect this energy, he needs to make sure that he couples to the field at the right time, and does not miss the sender's lightrays.

Whereas the energy density operator at a given spacetime point can be expressed in terms of field operators located on the boundary of its past lightcone, the decomposition of the field amplitude into operators acting on the field at earlier times involves operators inside the entire past lightcone.
Hence the field amplitude at a given point contains information from its entire past lightcone. This can be used to transmit information with signaling devices that  suitably couple to the field amplitude, even when the receiver misses the sender's lightrays.

The massless Klein-Gordon equation in (1+1)-dimensional Minkowski space reads \cite{birrell_quantum_1982}
\begin{align}
\left(\frac{\partial^2}{\partial t^2}-\frac{\partial^2}{\partial x^2}\right) \phi(t,x)=0
\end{align}
and we may expand the field operator in terms of plane wave field modes
\begin{align}
\phi(x,t)= \integral{k}{-\infty}\infty \frac1{\sqrt{4\pi|k|}} \left(\ee{-\ii (|k|t-kx)} a_k+ \ee{\ii (|k|t-kx)}a^\dagger_k\right).
\end{align}
The collection of all field operators $\phi(x,t)$ and their canonical conjugate momentum $\pi(t,x)=\partial_t\phi(t,x)$ located on a Cauchy hypersurface forms a complete set of observables \cite{wald_quantum_1994}. This means that any observable acting on the field can be expressed in terms of field operators $\left.\phi(t,x)\right|_{t=t^*}$ and $\pi(t,x)|_{t=t^*}$ located only on an arbitrarily chosen slice of constant time $t=t^*$.
In particular we have \cite{hotta_quantum_2008}:
\begin{align}
\phi(t,x)&=\frac12\left(\phi(0,x+t)+\phi(0,x-t)+\integral{y}{x-t}{x+t}\pi(0,y) \right) \label{eq:phiformula} \\
\pi(t,x)&=\frac12 \left( \pi(0,x+t)+\pi(0,x-t)\right.\nonumber\\
&\qquad\qquad \left. +\partial_x\phi(0,x+t)-\partial_x\phi(0,x-t)\right).\label{eq:piformula}
\end{align}
Note that these formulae are time translation invariant, and thus for $s<t$
\begin{align}
\phi(t,x)&=\frac12\left(\phi(s,x+t-s)+\phi(s,x-t+s) \vphantom{\integral{y}{x-t+s}{x+t-s}}\right. \nonumber\\
&\qquad\qquad\left.+\integral{y}{x-t+s}{x+t-s}\pi(s,y) \right)
\end{align}
and correspondingly for $\pi(t,x)$. 
We see right from this formula, that the field amplitude operator $\phi(t,x)$ is equal to a composition of field operators on earlier time slices covering every point in the past lightcone. In contrast, the conjugate $\pi(t,x)$ is composed of operators located only on the boundary of the lightcone.

The energy density of the field, i.e. the $T_{00}$-component of the energy-momentum tensor, reads \cite{birrell_quantum_1982}
\begin{align}
T_{00}(t,x)&= \frac12\left( (\partial_t \phi(t,x))^2 +(\partial_x\phi(t,x))^2\right)\nonumber\\
&=\frac12\left( (\pi(t,x)^2+(\partial_x \phi(t,x))^2\right).
\end{align}
From equation \eqref{eq:phiformula} above we have
\begin{align}
\partial_x\phi(t,x)&=\frac12\left( \partial_x \phi(0,x+t)+\partial_x\phi(0,x-t)\right.\nonumber\\
&\qquad \quad\left.+\pi(0,x+t)-\pi(0,x-t)\right),
\end{align}
together with \eqref{eq:piformula} this gives
\begin{align}\label{eq:T00formula}
T_{00}(t,x)&=\frac14 \left( \left(\pi(0,x+t)+\partial_x\phi(0,x+t)\right)^2 
\right. \nonumber\\ &\qquad\quad\left.
+\left(\pi(0,x-t)-\partial_x\phi(0,x-t)\right)^2\right).
\end{align}
The energy density of the field at any point in spacetime can thus be written in terms of field operators located only on the boundary of the point's lightcone. Thus the energy density measured in the field at one point in spacetime could only have been detected  by earlier measurements on the boundary of the point's lightcone. Conversely, any energy injected into the field at some location will only be observable along the boundary of the future lightcone of the point.
This means that no energy can be transmitted through the field between timelike separated points in (1+1)-dimensional Minkowski space.

This can also be understood from separating the field into a left-moving and a right-moving sector.
The two terms on the right hand side of equation \eqref{eq:T00formula}, which sum up to give the energy density, have a direct interpretation in terms of the energy flux from the left- and right-moving sectors of the field.

The field operator can be split up into a left- and a right moving part
\begin{align}
\phi(t,x)=\phi_+(t,x)+\phi_-(t,x)
\end{align}
with
\begin{align}
\phi_\pm(t,x)=\integral{k}0\infty \frac{1}{\sqrt{4\pi k}} \left(\ee{-\ii k(t\pm x)} a_{k,\pm}+\ee{\ii k(t\pm x)} a^\dagger_{k,\pm}\right).
\end{align}
The conjugate momentum is correspondingly split  into
\begin{align}
\pi(t,x)=\pi_+(t,x)+\pi_-(t,x)\end{align}
with
$
\pi_\pm(t,x)=\partial_t\phi_\pm(t,x).
$
These operators depend only on the so called lightcone coordinates $t\pm x$, e.g., $\phi_-(t,x)=\phi_-(u,y)$ if and only if $t- x=u- y$.

Whereas operators acting on different sectors of the field always commute, for the operators acting on the same sector we have
\begin{align}
\comm{\phi_\pm(t,x)}{\phi_\pm(u,y)}&=\frac\ii4 \text{sgn}((u-t)\pm(y-x))\\
\comm{\phi_\pm(t,x)}{\pi_\pm(u,y)}&=\frac\ii2\delta((u-t)\pm(y-x))\\
\comm{\pi_\pm(t,x)}{\pi_\pm(u,y)}&=-\frac\ii2\delta'((u-t)\pm(y-x)) \label{eq:picomm}
\end{align}
where $\delta'(x)$ denotes the derivate of the Dirac delta distribution. The commutator between two different field amplitude operators acting on the same sector is always non-vanishing. 

This implies that in  models which allow observers to directly couple to the $\phi_\pm(t,x)$ operators, signaling would be possible between arbitrary points in spacetime.
This problem does not occur for observers that only couple to the canonically conjugate operators. In fact, even if just one of two parties couples only to the operators $\pi_\pm(t,x)$ as, e.g., typically assumed in quantum energy teleportation protocols \cite{hotta_quantum_2014,hotta_quantum_2008,hotta_controlled_2010,verdon-akzam_asymptotically_2016}, signaling is possible only between lightlike separated points in spacetime due to the singular support on the lightcone's boundary of commutators involving $\pi_\pm(t,x)$.

However, the separation of the field into left- and right-moving modes is still useful to gain insight into the field's characteristics: From the  definitions above follows
\begin{align}
\partial_x \phi_\pm(t,x)=\pm \partial_t\phi_\pm(t,x)=\pm\pi_\pm(t,x).
\end{align}
So $ \pi(t,x)\pm \partial_x\phi(t,x) = 2\pi_\pm(t,x)$, and
\begin{align}
T_{00}(t,x)=\left(\pi_+(t,x)\right)^2 +\left(\pi_-(t,x)\right)^2.
\end{align}
Which shows that the energy density of a massless Klein-Gordon field is the sum of the left and right moving energy flux.

\section{Energy injected to the vacuum by an Unruh-DeWitt detector}\label{sec:energyfromdet}

In this section we discuss how the interaction with a simple two-level system, namely an Unruh-DeWitt detector, changes the energy, and energy density of the vacuum state of the field. This illustrates how the general findings  from the previous section on the propagation of energy  apply to the situation, in which timelike, energyless signals were first discussed in \cite{jonsson_information_2015}.

There is a large body of literature treating the radiation and energy flux originating from a Unruh-DeWitt detector, in particular  on the question whether an accelerated detector emits radiation while the detector experiences the Unruh effect. It has been considered for stationary, and non-stationary couplings both of two-level, and harmonic oscillator type detectors in (1+1)- and (1+3)-dimensional Minkowski space. Later works on this topic which give a review previous work include \cite{lin_accelerated_2006,kim_quantum_1999,kim_radiation_1997,massar_vacuum_1996,audretsch_radiation_1994}.

Here, we calculate the leading order perturbative contributions to $\exptval{\normord{T_{tt}(t,x)}}$, the normal-ordered Hamiltonian density operator of the massless scalar field in (1+1)-dimensional Minkowski space, caused by a two-level Unruh-DeWitt detector which couples to field for a finite time.
This is the particular scenario in which timelike and energyless signals were  analyzed originally in \cite{jonsson_information_2015}.

In its simplest form the Unruh-DeWitt particle detector consists of a two-level system $\left\{\ket{e},\ket{g}\right\}$ with Hamiltonian $H_d=\Omega \ketbra{e}{e}$, which we assume to be in the pure state
\begin{align}
\ket{\psi_0}=\alpha \ket{e}+\beta \ket{g}
\end{align}
before it interacts with the field. It couples to the field via the common interaction Hamiltonian for pointlike Unruh-DeWitt detectors. In the Dirac interaction picture it reads
\begin{align}\label{eq:intHMink}
\intH(\tau)= \lambda \Omega \chi( \tau) \left( \ee{-\ii \Omega \tau} \ketbra{g}{e} + \ee{\ii \Omega \tau} \ketbra{e}{g}\right) \otimes \phi (t_a,x_a)
\end{align}
where $(t_a,x_a)=(t_a(\tau),x_a(\tau))$ is the detector's location in (1+1)-dimensional Minkowski spacetime at detector proper time $ \tau$, and $\lambda$ is a dimensionless coupling constant. The switching function $\chi(\tau)$ controls at which times the detector is coupling to the field. It takes real values, i.e., $\chi(\tau)\in[0,1]\subset\mathbb{R}$, and is assumed to be compactly supported.

We assume the field to start out in the vacuum state. Hence the initial state of field and detector is given by the density operator
\begin{align}
\rho_0=\ketbra{\psi_0}{\psi_0}\otimes\ketbra00.
\end{align}
Using time-dependent perturbation theory, the final state of field and detector after the detector has coupled to the field can be expanded as
\begin{align}
\rho &\sim \rho_0 + \underbrace{\left( U^{(1)} \rho_0+ \rho_0 U^{(1)\dagger} \right)}_{=:\rho^{(1)}} \nonumber\\
& \quad+ \underbrace{\left(  U^{(2)} \rho_0+\rho_0 U^{(2)\dagger} +U^{(1)} \rho_0 U^{(1)\dagger} \right)}_{=:\rho^{(2)}}+\mathcal{O}(\lambda^3)
\end{align}
where $\rho^{(1)}\sim\mathcal{O}(\lambda)$, and $\rho^{(2)}\sim\mathcal{O}(\lambda^2)$, and
\begin{align}
U^{(1)}&=-\ii\integral{\tau}{}{}\intH(\tau) \\
U^{(2)}&=-\integral{\tau}{}{}\integral{\tau'}{}{\tau} \intH(\tau) \intH(\tau').
\end{align}

The normal-ordered energy density of the field, expanded in terms of field mode operators reads
\begin{align}\label{eq:normordhamiltondensity}
&\normord{T_{tt}(t,x)}\nonumber\\
&=\integral{k}{-\infty}{\infty} \integral{k'}{-\infty}{\infty} \frac{\sqrt{|kk'|}}{8\pi}\left(1+{kk'}\right) 
\left( 2 \ee{\ii (k-k')^\mu x_\mu} a^\dagger_{k} a_{k'}
\right.\nonumber\\&\qquad\quad\left.
-\ee{-\ii (k+k')^\mu x_\mu} a_{k} a_{k'} -\ee{\ii (k+k')^\mu x_\mu} a^\dagger_{k} a^\dagger_{k'} \right).
\end{align}
with $k^\mu x_\mu=|k|t-k x$.
Its expectation value after the detector has coupled to the field has the expansion
\begin{align}
\exptval{\normord{T_{tt}(t,x)}} &\sim  \tr\left( \normord{T_{tt}(t,x)} \rho^{(1)}\right) \nonumber\\
&+ \tr\left( \normord{T_{tt}(t,x)} \rho^{(2)}\right)+\mathcal{O}(\lambda^3).
\end{align}
However, the first contribution vanishes, 
\begin{align}
\tr\left( \normord{T_{tt}(t,x)} \rho^{(1)}\right)=0,
\end{align}
and thus the leading order contribution to the field energy density is of order $\mathcal{O}(\lambda^2)$.

The leading order contribution to the normal-ordered energy density is found to be
\begin{align}\label{eq:leadingTtt}
&\frac1{\lambda^2 \Omega^2}\tr\left( \normord{T_{tt}(t,x)} \rho^{(2)}\right) \nonumber\\
&=  \integral{ \tau}{}{} \integral{ \tau'}{}{} \frac{\chi( \tau) \chi(\tau')}{4} \left( | \alpha|^2 \ee{\ii \Omega ( \tau- \tau')} + | \beta|^2 \ee{-\ii \Omega (\tau- \tau')}\right)
\nonumber\\&\qquad\,\,\,
\left(\delta(x_+-a_+)\delta(x_+-a'_+)+\delta(x_--a_-)\delta(x_--a'_-)\right)  \nonumber\\
&\,\,\,\,+ \left(  | \alpha|^2-| \beta|^2 \right)   \integral{ \tau}{}{} \integral{ \tau'}{}{\tau} \frac{\chi( \tau) \chi(\tau')}{2\pi} \sin( \Omega( \tau- \tau') )
\nonumber\\&\,\,\,\,\,
\left(\delta(x_+-a_+) \pv \frac1{x_+-a'_+}+\delta(x_--a_-) \pv \frac1{x_- -a'_-}\right)
\end{align}
where $x_\pm=t\pm x$, $a'_\pm=t_a(\tau')\pm x_a(\tau')$ and $a_\pm$ is defined accordingly.
At this point we have not yet introduced any assumptions on the detector's wordline, or the switching function. However, due to the $\delta(x_\pm-a_\pm)$-distributions, a non-zero contribution to the energy density of the field can only be found at points $(t,x)$ that have one lightcone coordinate, $t\pm x$, in common with a spacetime point at which the detector interacted with the field.
This means that the detector only affects the (average) energy density of the field at spacetime points which are lightlike connected to its worldline: Any energy the detector injects into the field propagates lightlike.

\begin{figure}
\centering
\includegraphics[width=\columnwidth]{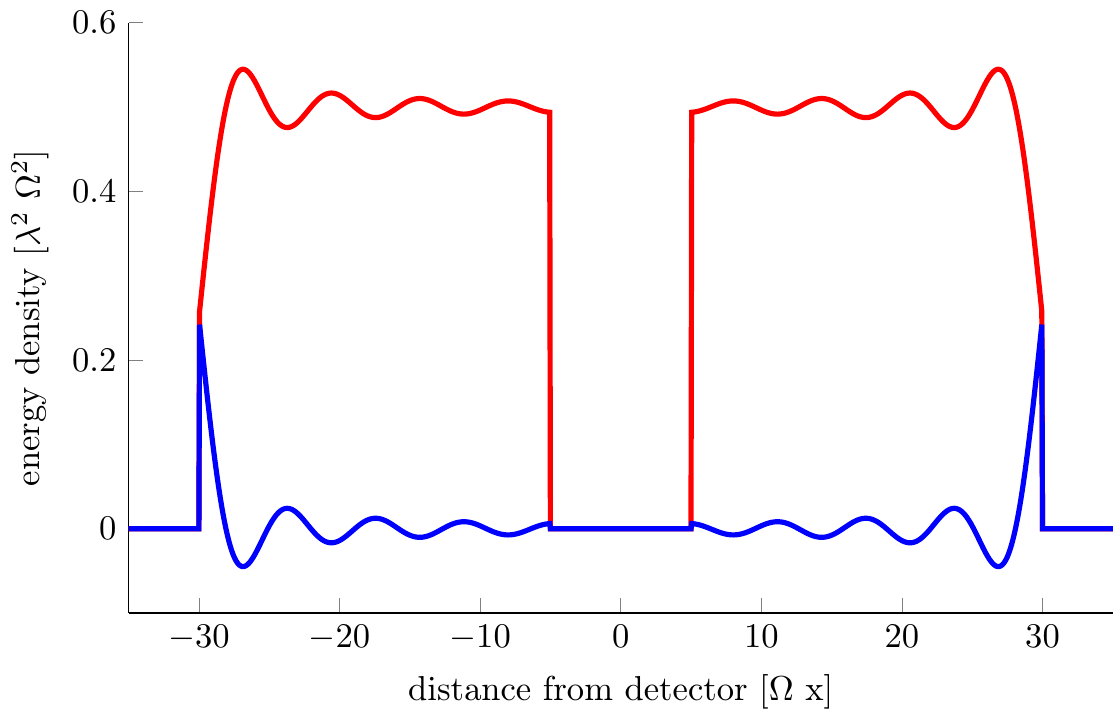}
  \caption{(Color online.) 
Leading order contribution to the normal ordered energy density $\exptval{\normord{T_{tt}(t ,x)}}$ from equation \eqref{eq:densityrestingdet} injected into the field by an Unruh-DeWitt detector at rest, starting out in its excited state (upper, red line), or its ground state (lower, blue line).
 The detector frequency $ \Omega$ is used to obtain dimensionless distances and times. The detector is coupled to the field at $x_a=0$ for the time interval  $t=0...25/ \Omega$. The graph shows the energy density on the spatial slice of 1+1D Minkowski space with $t=30/ \Omega$, i.e., the graph plots $\exptval{\normord{T_{tt}(t=30/ \Omega,x)}}$. The energy density is only non-vanishing at points which are ligthlike connected to the interacting detector.
   }
  \label{fig:energydensity}
\end{figure}

For a detector at rest, say at $x_a=0$, the detector's proper time coincides with coordinate time. If we couple the detector to the field for a time $T$ using a sharp switching function
\begin{align}
\chi(\tau)=\begin{cases} 1 \quad \text{ if } 0<\tau<T \\ 0 \quad \text{ else } \end{cases}
\end{align}
the leading order contribution to the energy density from equation \eqref{eq:leadingTtt} evaluates to
\begin{align}\label{eq:densityrestingdet}
&\exptval{\normord{T_{tt}(t,x)}}\sim\tr\left( \normord{T_{tt}(t,x)} \rho^{(2)}\right)\nonumber\\
&\qquad\sim \lambda^2 \Omega^2 \left( \frac{\chi(x_+) }{4} +\left( | \alpha|^2 - | \beta|^2\right) \frac{ \chi(x_+)}{2\pi} \text{Si}( \Omega x_+) \right.\nonumber\\
&\qquad\qquad\left.+\left(  \frac{\chi(x_-) }{4} + \left( | \alpha|^2 - | \beta|^2\right) \frac{ \chi(x_-)}{2\pi} \text{Si}( \Omega x_-)\right) \right).
\end{align}
A plot of the spatial profile of this contribution to the energy density is given in Fig.\@ \ref{fig:energydensity} for the case of a detector starting out in its excited state, $|\alpha|^2=1$, and for the case of a detector starting out in its ground state, $|\beta|^2=1$. For other input states the energy density can take any value in between, depending on the value of $|\alpha|^2-|\beta|^2$.

The energy density injected into the field by the detector interaction oscillates with a larger amplitude at times right after the onset of the interaction. In the plot these are the points most distant from the origin, where the detector is located. Towards later times, the oscillations diminish and the energy density approaches a constant limit of $\frac{\lambda^2\Omega^2}2$ for the excited detector, and zero for the detector in the ground state. Interestingly, the detector in the ground state causes negative energy densities to occur.

The abrupt switching through step functions used above can be problematic, e.g., for the calculation of detector excitation probabilities, because it does not comply with the mathematically rigorous requirement of being a smooth test function \cite{louko_transition_2008,satz_then_2007}. The change of energy expectation values for smooth switching functions converges to the results obtained for abrupt switching, as the smooth switching functions approach step functions \cite{jonsson_information_2015}. The oscillating fringes in the energy density of Fig.\@ \ref{fig:energydensity} are a consequence of steep switching, and are smoothened out for switching functions with lower slopes.

\begin{figure}
\centering
\includegraphics[width=\columnwidth]{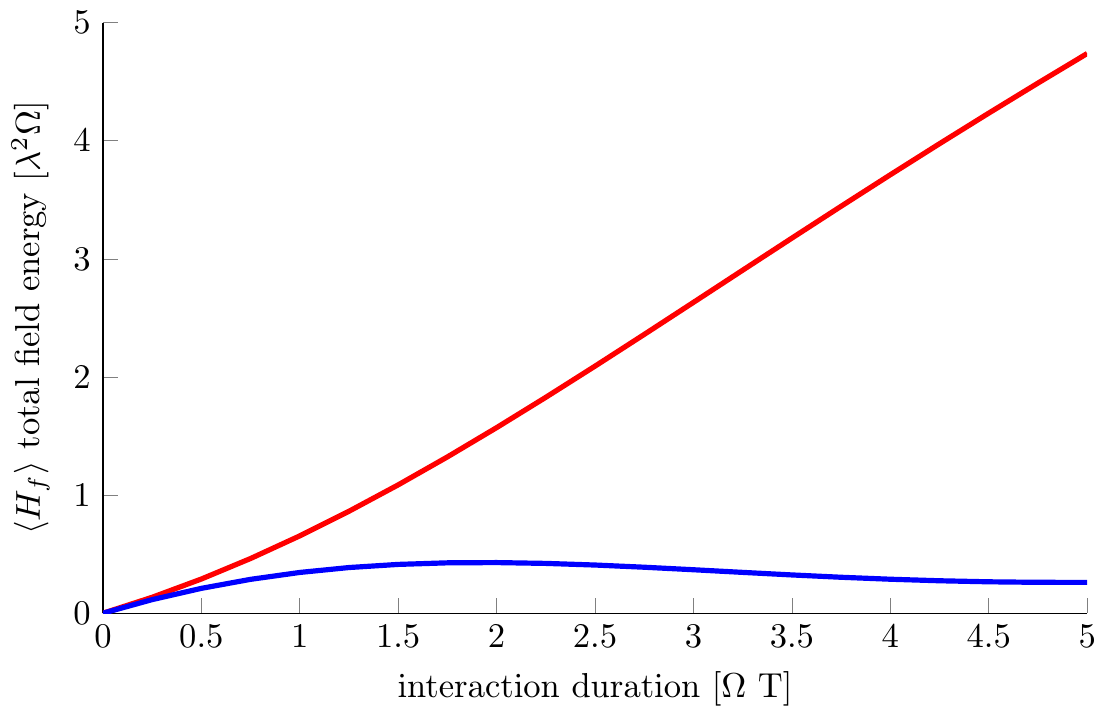}
  \caption{(Color online.) 
Leading order perturbative contribution to the total energy injected into the field by an Unruh-DeWitt detector, with energy gap $\Omega$, coupling to the field for the time interval $t=0...T$ while at rest, see \eqref{eq:totalenergyatrest}. The red (upper) line shows a detector starting out in its excited state. Here the contribution to the energy scales linearly with the total interaction time $ \Omega T$. The blue (lower) line shows a detector starting out in its ground state. Here the total energy approaches the limit of $\exptval{H_f}\rightarrow \lambda^2\Omega/\pi$ as $T\rightarrow\infty$, see \eqref{eq:betaenergylimit}. 
  }
  \label{fig:fieldenergy}
\end{figure}

The total energy injected into the field by the detector interaction can be obtained by integrating the expectation value of the energy density over a spatial slice of fixed time. This yields the expectation value of the field energy, because the Hamiltonian of the field, $H_f=\integral{x}{}{} \normord{T_{tt}(t,x)}$, is the integral over the energy density .
\begin{align}\label{eq:totalenergyatrest}
&\exptval{H_f}=\integral{x}{}{} \exptval{\normord{T_{tt}(t,x)}}\nonumber\\
&\sim   \integral{x}{}{} \tr\left( \normord{T_{tt}(t,x)} \rho^{(2)}\right)+\mathcal{O}(\lambda^3)\nonumber\\
&\sim {\lambda^2 \Omega^2 T} \!\left( \frac12\!+\!\frac{\left( | \alpha|^2 - | \beta|^2\right)}\pi\left( \text{Si}( \Omega T)+\frac{\cos( \Omega T)-1}{T \Omega}\right) \right)\nonumber\\
&\qquad\qquad  +\mathcal{O}(\lambda^3).
\end{align}
A plot of the total energy injected into the vacuum state of the field by a detector in its ground, or in its excited state is given in Fig.\@ \ref{fig:fieldenergy}.

When the detector starts out in its excited state, this contribution scales linearly with the total interaction time, so that $\exptval{H_f}{\sim} \lambda^2 \Omega^2 T$ as $\Omega T\rightarrow\infty$.
For long interaction times this divergence might be overcome by higher order perturbative contributions, or otherwise it indicates the limits of the perturbative regime.

For a detector starting in its ground state the limit of long interaction times yields
\begin{align}\label{eq:betaenergylimit}
\exptval{H_f}\sim\lambda^2\frac\Omega\pi
\end{align}
as $\Omega T\rightarrow\infty$.
This means that even in the limit of an infinitely long interaction time window, a detector starting out in its ground state injects energy into the field. This energy corresponds to the energy injected early on the interaction after the abrupt switch-on of the interaction. Any energy that is injected into the field propagates away at the speed of light though. Hence there are no means to retain it later at the detector's location.

In a certain sense this energy which is injected into the field by a detector in its ground state, can be viewed as a binding energy between the detector and the field, because this elevation in the field energy is accounted for by an elevated energetic cost of switching off the coupling between detector and field \cite{jonsson_information_2015}.

It is interesting to note that the leading order contribution to the field Hamiltonian can be obtained from
\begin{align}
\exptval{H_f}\sim \integral{x}{}{} \tr\left( \normord{T_{tt}(t,x)} U^{(1)} \rho_0 U^{(1)\dagger} \right)
\end{align}
alone, whereas all terms of $\rho^{(2)}$ are contributing to the energy density
\begin{align}
&\exptval{\normord{T_{tt}(t,x)}}\nonumber\\
&\sim \tr\left(\normord{T_{tt}(t,x)}\left(U^{(1)} \rho_0 U^{(1)\dagger}+ U^{(2)} \rho_0+\rho_0 U^{(2)\dagger } \right)\right).
\end{align}
This is because when integrated over a spatial slice of constant time, the latter terms average out
\begin{align}
\integral{x}{}{} \tr\left(\normord{T_{tt}(t,x)}\left( U^{(2)} \rho_0+\rho_0 U^{(2)\dagger} \right)\right)=0.
\end{align}
So the energy injected into the field by a detector on an arbitrary worldline is determined by the more symmetric term above alone, which yields
\begin{align}
&\integral{x}{}{}\tr\left( \normord{T_{tt}(t,x)} U^{(1)} \rho_0 U^{(1)\dagger} \right)\nonumber\\
&\quad= \lambda^2 \integral{k}{-\infty}\infty\integral{ \tau}{-\infty}\infty \integral{ \tau'}{-\infty}{\infty} \frac{\chi( \tau) \chi(\tau')}{2\pi}  
\nonumber\\ &\qquad\qquad
\ee{\ii k^\mu (a-a')_\mu} \left( | \alpha|^2 \ee{-\ii \Omega ( \tau- \tau')} + | \beta|^2 \ee{\ii \Omega (\tau- \tau')}\right).
\end{align}

The infrared ambiguity of the massless field in (1+1)-dimensional Minowski space does not affect the expectation values of the energy-momentum tensor. This explains why the results above for the detector's effect on the field energy, and energy density all are finite, whereas the excitation probability for a pointlike detector in 1+1 dimensions is infrared-divergent and has to be regularized \cite{takagi_vacuum_1986,juarez-aubry_onset_2014}. 

To confirm that signaling between timelike separated detectors is not an artefact of  infrared effects, in the following sections we study scenarios inside a cavity, where these problems do not appear, because of the discrete mode structure of the field.

\section{Timelike signaling without zero mode}\label{sec:cavity}
Inside a cavity, the continuous momentum integral is replaced by a discrete sum over the field modes, which gets rid of infrared divergences. In a periodic cavity, attention has to be paid to the treatment of the zero-mode of the field, whose importance in detector-field interactions was analysed in  \cite{martin-martinez_particle_2014}.

The constant value of the field commutator inside the future lightcone might raise the question, whether this feature is connected to the zero-mode. However, the following analysis of the field commutator inside a cavity with Dirichlet boundary conditions shows that timelike signals also occur in massless fields in  (1+1)-dimensional cavities in the absence of a zero-mode.

Imposing Dirichlet boundary conditions means that the field vanishes at the cavity boundaries for $x=0,L$. The field operator inside the cavity can be expanded as
\begin{align}\label{eq:fieldop2d}
\phi(t,x)= \sum_{j=1}^\infty \frac{1}{\sqrt{j \pi} } \sin(j \pi x/L) \left( a_j \ee{-\ii \frac{ j \pi }{L} t } + a^\dagger_j \ee{\ii \frac{j \pi }{L}t} \right)
\end{align}
without any zero mode. Using this expansion yields the following expression for the field commutator
\begin{align}
&\comm{ \phi(t_1,x_1)}{\phi(t_2,x_2)}  \nonumber\\ 
& =-  \sum_{j=1}^\infty  \frac{2\ii}{\pi j}  \sin(j \pi x_1/L) \sin(j \pi x_2/L) \sin\left(j \pi (t_1-t_2)/L\right)
\end{align}
which is analytically summable, and equal to
\begin{align}
&\comm{\phi(x_1,t_1)}{\phi(x_2,t_2)} \nonumber\\
&\quad= \frac{\ii}2 \left(\left\lfloor \frac{t_1-t_2+x_1-x_2}{2L} \right\rfloor+\left\lfloor \frac{t_1-t_2-x_1+x_2}{2L} \right\rfloor \right.\nonumber\\
&\qquad\left.-\left\lfloor \frac{t_1-t_2+x_1+x_2}{2L} \right\rfloor- \left\lfloor \frac{t_1-t_2-x_1-x_2}{2L} \right\rfloor \right)
\end{align}
where $\lfloor x\rfloor=\text{floor}(x)$ denotes the floor function.

\begin{figure}
\centering
\includegraphics[width=\columnwidth]{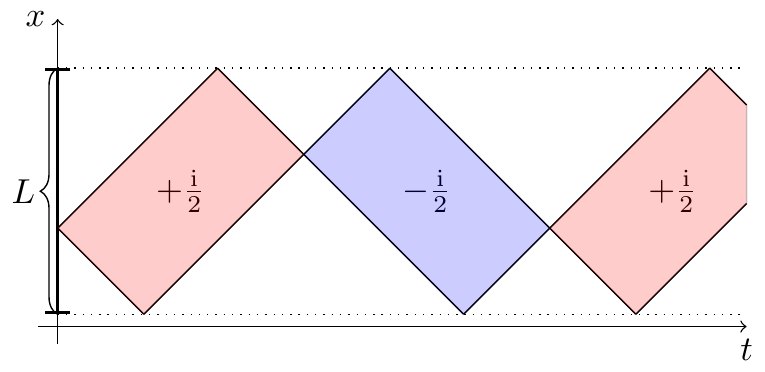}
\caption{(Color online) Sketch of  $\left[\phi(t_1=0,x_1),\,\phi(t,x)\right]$, the commutator of a massless Klein-Gordon field in a one-dimensional cavity of length $L$ with Dirichlet boundary conditions. In the red areas it takes the value $+\frac\ii2$ and in the blue areas it takes the value $-\frac\ii2$.
}
\label{fig:comm1}
\end{figure}

Figure \ref{fig:comm1} shows a sketch of the commutator  $\comm{ \phi(t_1,x_1)}{\phi(t_2,x_2)}$. When $(t_1,x_1)$ and $(t_2, x_2)$ are close, the commutator is identical to the commutator in (1+1)-dimensional Minkowski space: It vanishes at spacelike separations, and it takes the value $+\frac\ii2$ when $(t_1,x_2)$ lies in the future lightcone of $(t_1,x_1)$. For large separations, the lightcone structure is reflected by the cavity walls, such that the commutator is periodic in $t_2-t_1$ with periodicity $2L$.

This shows that signaling  to a receiver localized inside the first, unreflected portion of the sender's lightcone is comparable to the two parties being located in Minkowski space (The noise of the receiver will be different inside the cavity see \cite{jonsson_quantum_2014} and Appendix \ref{app:timelikemean}). In particular, timelike and energyless signaling does not rely on the presence of a zero-mode in the field.

\section{Timelike signaling beyond perturbation theory} \label{sec:honew}

The discussion of Section \ref{sec:propagation} shows that timelike signaling is possible between any type of signaling device that couples to the amplitude of the massless field. In particular, these timelike signals are always decoupled from the flow of energy that the sender may have injected into the field, because the field energy density propagates strictly at the speed of light.

However, there are many more questions about the properties of timelike signals to be explored in future research: How much quantum information can coherently be transmitted via timelike signals? To what extent does a single sender's timelike signal lead to correlations between different simultaneous receivers? These questions are difficult to address within perturbation theory. 

Therefore, in this section we  go beyond the perturbative treatment of two-level particle detectors in \cite{blasco_violation_2015,blasco_timelike_2016,jonsson_information_2015} and Section \ref{sec:energyfromdet}, and study timelike signaling, non-perturbatively, between harmonic oscillators coupling to the field inside a cavity. 
In this model the full unitary time evolution of the field and the signaling devices can be calculated using symplectic methods  \cite{brown_detectors_2013,bruschi_time_2013}, both for the quantum system as well as for classical oscillators coupling to a classical field.

\begin{figure}
\centering
\includegraphics[width=\columnwidth]{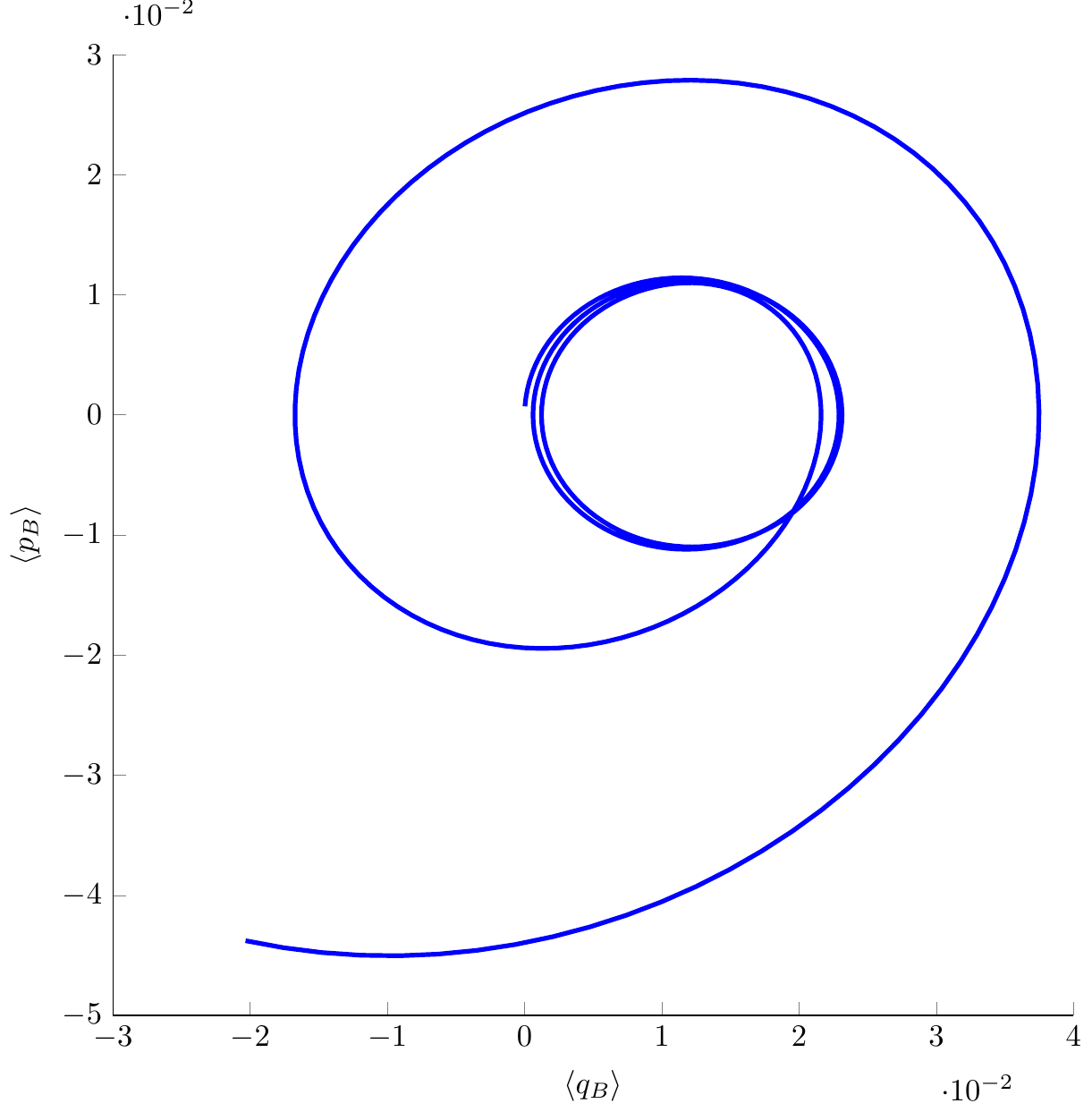}
  \caption{(Color online.) Phase space plot of receiver's final mean $\exptval{ \mathbf{x_B}(T_2)}=\left(\exptval{q_B},\exptval{p_B}\right)=\frac1{\sqrt{2}}\left(a_B+a_B^\dagger,a_B-a_B^\dagger\right)$. The inner circles correspond to timelike separation between the sender and the receiver (see Fig.\@ \ref{fig:timelike_mean_radial}). The sender is located at $a=0.5L$, and the receiver at $b=0.6L$. The sender couples to the field for $t=0...0.3L$. The receiver couples to the field for $t=0.46L... T_2$. The points in this plot correspond to $T_2=0.46...1.2L$. The detectors' energy gap is $ \Omega=10 \pi/L$ and the coupling constant $\lambda=0.075$. For the computation $N=200$ field modes were taken into account.
  }
  \label{fig:timelike_mean_phasespace}
\end{figure}

\begin{figure}
\centering
\includegraphics[width=\columnwidth]{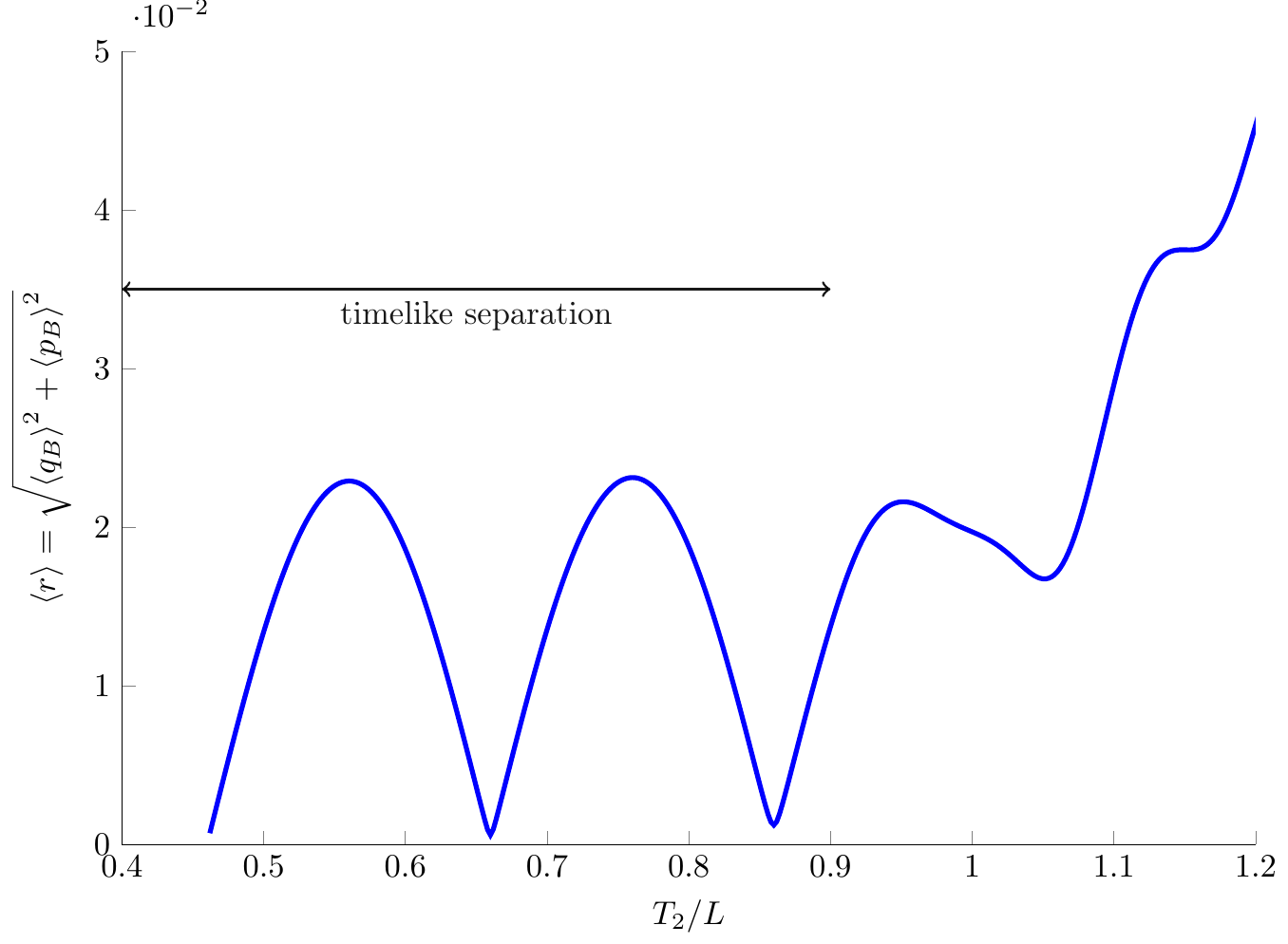}
  \caption{(Color online.)   Displacement  $\exptval{r}=\sqrt{\exptval{q_B}^2+\exptval{p_B}^2}$ of the receiver's final state over different coupling times for the receiver in the same setting as Fig.\@ \ref{fig:timelike_mean_phasespace}. The receiver and sender are strictly timelike separated  for $T_2<0.9L$. For lightlike separations between sender and receiver see Appendix  \ref{app:timelikemean}.
  }
  \label{fig:timelike_mean_radial}
\end{figure}

The setup we are discussing in this section differs from the one discussed earlier in Section \ref{sec:energyfromdet} in two points: 
Instead of a two-level system,  the particle detector now  consists of an harmonic oscillator. And instead of free Minkowski space, the field is now confined inside a cavity with vanishing Dirichlet boundary conditions, just as in the previous section.

The interaction Hamiltonian between a single harmonic oscillator detector and the field inside the cavity reads
\begin{align}\label{eq:HintHO}
H_{I,d}= \lambda \Omega \left( \ee{-\ii \Omega t} a_d + \ee{\ii\Omega t} a^\dagger_d\right) \phi(t,x_d)
\end{align}
in the interaction picture.
Where now the operators $a_d$ and $a^\dagger_d$ are the ladder operators acting on the harmonic oscillator detector, and $ \Omega$ is the harmonic oscillator's energy level spacing. As before, $\lambda$ is a dimensionless coupling constant. 
Here, in comparison to \eqref{eq:intHMink},  we also dropped the switching function, and assume instead that the interaction is switched on and off instantaneously.
In the signaling scenario that we study here, the interaction Hamiltonian will have two terms as in \eqref{eq:HintHO}, one for the sender ($d=A$) and one for the receiver ($d=B$).

The advantage of a Dirichlet cavity is that  the full unitary time evolution of the detectors and the field can be calculated numerically with the methods developed in \cite{brown_detectors_2013}. In contrast, in a periodic cavity the zero mode of the field would require separate analysis. 

However, this advantage comes  at the price of losing spatial translational invariance: 
Inside a Dirichlet cavity the field modes have varying intensity at different locations, because they exhibit maxima and node points due to the boundary conditions. These maxima and nodes lie at different points for the individual modes depending on their wave length. Therefore also the behaviour of a detector coupling to the field inside a Dirichlet cavity depends on the detector's location.

As far as signaling between two detectors is concerned, these effects are negligible when sender and receiver are strictly timelike separated, in the sense that not even lightrays reflected by the cavity walls can connect the sender to the receiver. App. \ref{app:timelikemean} discusses this in detail.

In a numerical calculation, of course, only a finite number of $N$ field modes can be taken into account. This means introducing a UV cutoff in the field expansion \eqref{eq:fieldop2d}
\begin{align}
\phi(t,x)= \sum_{j=1}^N \frac{1}{\sqrt{j \pi} } \sin(j \pi x/L) \left( a_j \ee{-\ii \frac{ j \pi }{L} t } + a^\dagger_j \ee{\ii \frac{j \pi }{L}t} \right).
\end{align}
This cutoff needs to be   large enough to capture the relativistic properties of the field with accuracy \cite{brown_detectors_2013,jonsson_quantum_2014}.


The interaction Hamiltonian  \eqref{eq:HintHO} between field and detectors is strictly quadratic in the creation and annihilation operators, without any linear terms. 
Under the time evolution of such a Hamiltonian  Gaussian states remain Gaussian. Therefore, as explained in detail in \cite{brown_detectors_2013}, the detectors' and field's combined time evolution can be conveniently described in the symplectic formalism, i.e., through the total system's symplectic matrix $\mathbf{S}(t)$.

To this end, we organize the quadrature operators of the two detectors (whose annihilation operators we denote as $a_A$ and $a_B$) and the $N$ field modes  into one $(2N+4)$-dimensional vector
\begin{align}
\mathbf{x}= \frac1{\sqrt2} &\left(  \left(a_A+a_A^\dagger\right),\left(a_B+a_B^\dagger\right), ..., \left( a_N+a_N^\dagger \right), \right.\nonumber\\
 &\left.\qquad\left(a_A-a_A^\dagger\right),\left(a_B-a_B^\dagger\right), ..., \left( a_N-a_N^\dagger \right) \right),
\end{align}
such that the first $(N+2)$ entries correspond to the detectors' and field modes' canonical position operators, and the second $(N+2)$ entries correspond to the canonical momentum operators. 

A Gaussian state is fully characterized by its first two moments \cite{adesso_continuous_2014,braunstein_quantum_2005,weedbrook_gaussian_2012}, i.e., by the expectation value of the quadrature operator vector $\exptval{\mathbf{x}}$, and by its covariance matrix $\mathbf{ \sigma}$. We use the convention 
\begin{align}\label{eq:covariancemat}
{\sigma}_{i,j}\equiv \frac12 \exptval{\{ \Delta \mathbf{x}_i, \Delta \mathbf{x}_j \}}=\frac12 \left( \exptval{\mathbf{x}_i \mathbf{x}_j+\mathbf{x}_j \mathbf{x}_i} - 2 \exptval{\mathbf{x}_i}\exptval{\mathbf{x}_j} \right).
\end{align}
such that the ground state of a harmonic oscillator has the covariance matrix $ { \sigma}_{i,j} =\frac12 \delta_{ij}$.

The time evolution of the state's moments are obtained by multiplication with the symplectic matrix,
\begin{align}
&\exptval{\mathbf{x}(t)}=\mathbf{S} \exptval{\mathbf{x}_0} \label{eq:meandisplacement}\\
&\mathbf{ \sigma}(t)= \mathbf{S} \mathbf{ \sigma}_0 \left(\mathbf{S}\right)^{T} \label{eq:covariancedisplacement}
\end{align}
where $ \mathbf{S}^T$ denotes the transpose of the symplectic matrix. 

Note that by the Ehrenfest theorem the first moment, the mean $\exptval{ \mathbf{x}}$,  evolves according to the classical equations of motion. 
Therefore the  behaviour we observe for the mean of a Gaussian state also arises in a classical system of harmonic oscillators coupling to a classical field.

We will now study the following signaling scenario: Initially the field is in its vacuum state, i.e, all the field modes are in their ground state. 
The sender's detector is initialized in some pure Gaussian state, and the receiver is initialized in its ground state.
First the sender's detector is coupled to the field for some time $t=0...T_A$. Then , with some delay, the receiver is coupled to the field for some time $t=T_1...T_2$. 
The delay between the two couplings $T_1-T_A$ is chosen such that the sender and receiver are strictly timelike separated, i.e., no light ray can connect them while they are coupling to the field.

After the receiver is decoupled from the field, the mean displacement of the sender's harmonic oscillator depends linearly on the initial mean displacement of the sender. Because from equation \eqref{eq:meandisplacement} follows
\begin{align}\label{eq:matrixeqnmean}
\exptval{ \mathbf{x_B}(T_2)} =
\begin{pmatrix} S_{2,1} & S_{2, N+3}\\ S_{N+4, 1} & S_{N+4,N+3}\end{pmatrix}
\exptval{ \mathbf{x_A}(0)}
\end{align}
with $ \mathbf{x_d}=\left(q_d,p_d\right)=\frac1{\sqrt 2}\left( a_d+a_d^\dagger, \, a_d-a_d^\dagger\right)$ for  $d=A,B$. Here $S_{i,j}$ denote the matrix elements of the symplectic matrix $\mathbf{S}=\left( S_{i,j}\right)$.

Figures \ref{fig:timelike_mean_phasespace} and \ref{fig:timelike_mean_radial} show that even for strictly timelike separation between sender and receiver, the sender can induce a displacement of the receiver's final state. For this the sender needs to prepare an initial state with non-vanishing displacement in canonical momentum $\exptval{p_A}$, such as $\exptval{ \mathbf{x_A}(0)}= (0,1)$. 
When the sender interacts with the field for times on the order of a few detector periods $ \Omega/(2 \pi)$, and is resonant with higher modes of the cavity, an initial displacement in canonical position $\exptval{q_A}$, hardly affects a timelike separated receiver. It only affects receivers reached by (reflected) light rays, as seen in App.\@ \ref{app:timelikemean}. This is different for detectors resonant with the base mode of the cavity ($ \Omega= \pi/L$) where also $\exptval{ \mathbf{x_A}}=(1,0)$ leads to timelike signaling, as seen in App.\@ \ref{app:basemode}. 

To maximize timelike signaling effects, the sender needs to couple to the field for a multiple plus a half of a detector period, i.e., for $T_A=\left(2n+1\right) \Omega/(4\pi)$ with $n\in\mathbb{N}$. The timelike effects vanish when the sender is coupled for an integer multiple of its detector period, i.e., for $T_A= n \Omega/ (2\pi)$.

In the  example of Figure \ref{fig:timelike_mean_phasespace} and \ref{fig:timelike_mean_radial} (coupling constant $ \lambda=0.075$) we observe  a displacement of the receiver's final state which is  in the percentile range of the sender's initial displacement. (Higher values of the coupling constant increase the effect.) 
This displacement decreases the overlap between the sender's final state and its initial ground state and could thus be used classical information transmission \cite{jonsson_information_2015}, both via the quantum field as well as in its classical analogue.

When the sender is initialized in a zero-mean Gaussian state (just as the receiver and the field modes always are as they are initialized in their ground state), then no displacement of the receiver's final state occurs. Because under  Hamiltonians that are only quadratic in creation and annihilation operators zero-mean Gaussian remain zero-mean Gaussian states.

However, timelike signals can be evoked not only by the sender's initial mean $\exptval{ \mathbf{x_A}(0)}$, but also by its second moments. 
To analyze the interdependence of the sender's and the receiver's second moments it is convenient to introduce a three-vector notation for the dectectors' covariance matrices. The covariance matrix of a single detector is a symmetric 2x2-matrix, i.e., it only has three independent elements because the off-diagonal elements are identical. We denote the entries of the sender's covariance matrix by
\begin{align}
\vec{ \mathbf{ \sigma_A}}=\frac12 \begin{pmatrix} \exptval{\{ \Delta q_A, \Delta q_A \}} \\ \exptval{\{ \Delta p_A, \Delta p_A \}}\\ \exptval{\{ \Delta q_A, \Delta p_A \}} \end{pmatrix}
= \begin{pmatrix} \sigma_{1, 1}\\ \sigma_{N+3,N+3} \\ \sigma_{1, N+3}\end{pmatrix}
\end{align}
where $\sigma_{i,j}$ refers to the matrix elements of the total system's covariance matrix as defined in \eqref{eq:covariancemat}. Accordingly, for the sender we define
\begin{align}
\vec{ \mathbf{ \sigma_B}}=\frac12 \begin{pmatrix} \exptval{\{ \Delta q_B, \Delta q_B \}} \\ \exptval{\{ \Delta p_B, \Delta p_B \}}\\ \exptval{\{ \Delta q_B, \Delta p_B \}} \end{pmatrix}
= \begin{pmatrix} \sigma_{2, 2}\\ \sigma_{N+4,N+4} \\ \sigma_{2, N+4}\end{pmatrix}.
\end{align}
It then follows from equation \eqref{eq:covariancedisplacement} that the receiver's final covariance elements are given by
\begin{widetext}
\begin{align}\label{eq:matrixeqnCM}
\vec{\mathbf{ \sigma_B}}(t) &= \begin{pmatrix} (S_{2,1})^2 & (S_{2,N+3})^2 & 2S_{2,1} S_{2,N+3} \\ (S_{N+4,1})^2 & (S_{N+4,N+3})^2 & 2 S_{N+4,1}  S_{N+4,N+3} \\ S_{2,1} S_{N+4,1} & S_{2,N+3} S_{N+4,N+3} & S_{2,1} S_{N+4,N+3} + S_{2,N+3} S_{N+4,1} \end{pmatrix} \vec{\mathbf{ \sigma_A}}(0) + \sum_{i\neq1,N+3}\frac12 \begin{pmatrix}  (S_{2,i})^2 \\ (S_{N+4,i})^2 \\ S_{2,i} S_{N+4,i}\end{pmatrix}
\end{align}
\end{widetext}
where we used that all field modes and the receiver's detector are initially in their ground state. 
The covariance matrix elements of the receiver's final state consist of an affine part, which can be viewed as background noise, on top of which a contribution is added which is linear in the sender's initial covariance matrix elements.

Due to these contributions there is a certain excitation probability $P_e$ to measure the receiver in a state other than the ground state after its interaction with the field. It can be calculated from the covariance matrix of the receiver's (zero-mean) Gaussian state by \cite{brown_detectors_2013}
\begin{align}
P_e=1-\frac2{\sqrt{4\det\sigma+2\tr\sigma+1}}.
\end{align}
If this probability is changed by the sender's action, this can be used to transmit classical information \cite{jonsson_information_2015}.

\begin{figure}
\centering
\includegraphics[width=\columnwidth]{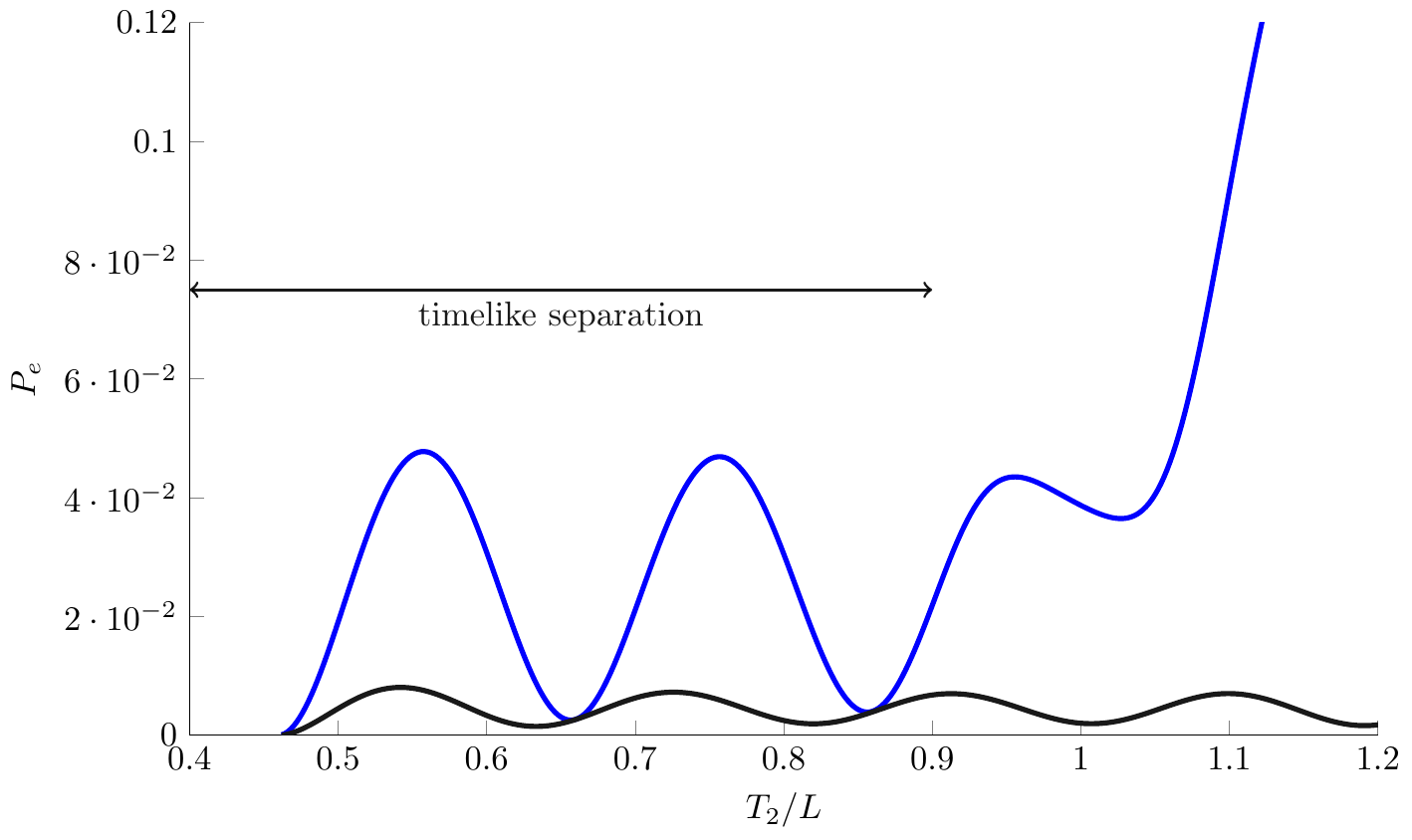}
  \caption{(Color online.) 
  Timelike signaling via thermal states. $P_e$ is the excitation probability for the receiver.   
   The lower (black) line shows $P_e$ when the receiver is coupling to the vacuum state of the field, in the absence of a sender. The upper (blue) line shows the elevation of $P_e$ by the sender in the signaling scenario.
  The sender is initialized in a thermal state with $ \Omega/ \tau = 6\cdot10^{-3}$, located at $a=0.5L$, and coupled to the field for $t=0...0.3L$. The receiver is located at $b=0.6L$, and coupled to the field for $t=0.46L...T_2$. Hence sender and receiver are strictly timelike separated for $T_2<0.9L$. The detectors' energy spacing is $ \Omega=10 \pi/L$, the coupling constant $ \lambda=0.075$. For the numerical calculations $N=200$ modes were used.
  }
  \label{fig:thermaltimelike}
\end{figure}

Figure \ref{fig:thermaltimelike} shows an example where initializing the sender in a thermal state, as an example of a zero-mean state, can be used for signaling to a timelike separated receiver.  
Such a thermal state of the sender's detector of temperature $ \tau$ has the covariance matrix
\begin{align}
\mathbf{ \sigma}_{th}= \frac12\coth\left( \frac{ \Omega}{2 \tau}\right)  \mathbb{I}. 
\end{align}

Just as for the effects on the mean displacement, the sender needs to couple for a time $T_A=(2n+1) \Omega/ (4\pi)$ to maximize the effect on a strictly timelike separated receiver's covariance matrix. These effects vanish when the sender is coupled for $T_A=n \Omega/(2 \pi)$, i.e., for a multiple of a detector period.

Similarly, it is the variance of the sender's initial canonical momentum $ \sigma_{N+3\,N+3}= \Delta p_A= \exptval{p_A^2}-\exptval{p_A}^2$ which evokes stronger timelike signaling effects than the sender's initial variance in canonical position $ \sigma_{1\,1}= \Delta x_A=\exptval{x_A^2}- \exptval{x_A}^2$, as a comparison of 
equations \eqref{eq:matrixeqnmean} and  \eqref{eq:matrixeqnCM} already suggests. 
This can be demonstrated by initializing the sender in differently rotated squeezed states, where we find that timelike signaling is stronger for states squeezed in position (with large $ \Delta p_A$), while  timelike signaling effects are tiny for momentum squeezed states (with large $ \Delta x_A$). 
Although when sender and receiver are connected by (reflected) lightrays the signals from  momentum squeezed states are comparable in size to the signals from position squeezed states. (See App.\@ \ref{app:timelikemean}.)

The results of this section show that signaling is possible between strictly timelike separated harmonic oscillators coupling to the field inside a cavity, and that these signals can be transmitted both vie the first as well as via the second moments of Gaussian states. Most importantly, these signals were treated non-perturbatively using symplectic methods. 
This makes the model an interesting tool for future research into the classical and quantum information capacity of timelike and  energyless signals.

\section{Conclusions and Outlook}

The present results explain why in 1+1 spacetime dimensions,  and possibly in experimental setups that confine massless fields to effectively 1+1 dimensions,  energy transport and information transmission can be decoupled:  
It is generally possible to send information via the field to timelike separated points, because the amplitude of the field depends on the interior of its past lightcone.
This is impossible for energy, which can only be transported at the speed of light, because the energy density at any given spacetime point only depends on points on the lightcone's boundary.

Thus timelike, energy less signals, first observed in \cite{jonsson_information_2015}, are indeed a general phenomenon of massless fields   and are not  limited to a particular model of signaling device.

We demonstrated that the effect occurs in cavities with Dirichlet boundaries, which shows that  the effect is not dependent on  infrared divergences or on the zero-mode of the field.  
Further, the appearance of timelike signals in a non-perturbative treatments of harmonic oscillators interacting with the field inside a cavity demonstrates that the effect is not a mere artifact of perturbation theory either.

The observation of timelike signals in fields inside cavities strongly motivates investigating a possible experimental demonstration of this effect, e.g., in superconducting circuits.  
As we showed the effect is a general feature of massless fields in 1+1 dimensions, so it is expected to appear in a variety of systems. 
Nevertheless, its implementation certainly involves a number of challenges: 
To begin with, the cavity has to be long as compared to the total interaction time between detector system and the field, in order to allow for strict timelike separation between the detectors. Furthermore, the coupling between detector and field needs to be switchable on time scales short in comparison to the inverse of the detector's internal energy scale.

In addition to (1+1)-dimensional settings, also (2+1)-dimensional scenarios could be interesting both in view of experiments, as well from a theoretical perspective: In (2+1)-dimensional Minkowski space both the field, and its conjugate are non-vanishing inside the lightcone. This means that both information and energy can partially propagate slower than the speed of light. The commutators decay with different powers of the distance, though, which raises the question to what extent the flow of information overcomes the flow of energy quantitatively \cite{jonsson_decoupling_2016}.

This is also relevant in view of the question of how much quantum information can be carried by signals that do not transmit energy? The ability of timelike signals in 1+1 dimensions to fill the entire future lightcone of the sender, can be used to broadcast classical information to an arbitrary large number of  independent, identical receivers. However, this is impossible for quantum information which cannot be broadcast, because it cannot be cloned. Therefore, ultimately, the quantum capacity of energyless signals has to be limited. In this context it is interesting to note the recent result  that, conversely, the transport of energy between quantum systems requires quantum correlations in the form of discord \cite{lloyd_no_2015}. It should be an interesting question for future research to explore whether these observations could point towards a deeper link between the transport of energy and the transmission of quantum information.


The observation of timelike signaling between harmonic oscillators made above is also important for another reason: To this model the widely developed methods of continuous variable quantum information are applicable \cite{braunstein_quantum_2005,weedbrook_gaussian_2012} which adds an interesting perspective for ongoing and future research on this topic.

\section{Acknowledgements}
I am grateful to Achim Kempf and Eduardo Mart\'{i}n-Mart\'{i}nez for helpful discussions and feedback, and their continuing support.
I also thank Katja Ried and Guillaume Verdon-Akzam for helpful discussions.
And I gratefully acknowledge support by the German National Academic Foundation.

\appendix
\section{Influence of position inside Dirichlet cavity for harmonic oscillator detectors}\label{app:timelikemean}

\begin{figure}
\centering
\includegraphics[width=\columnwidth]{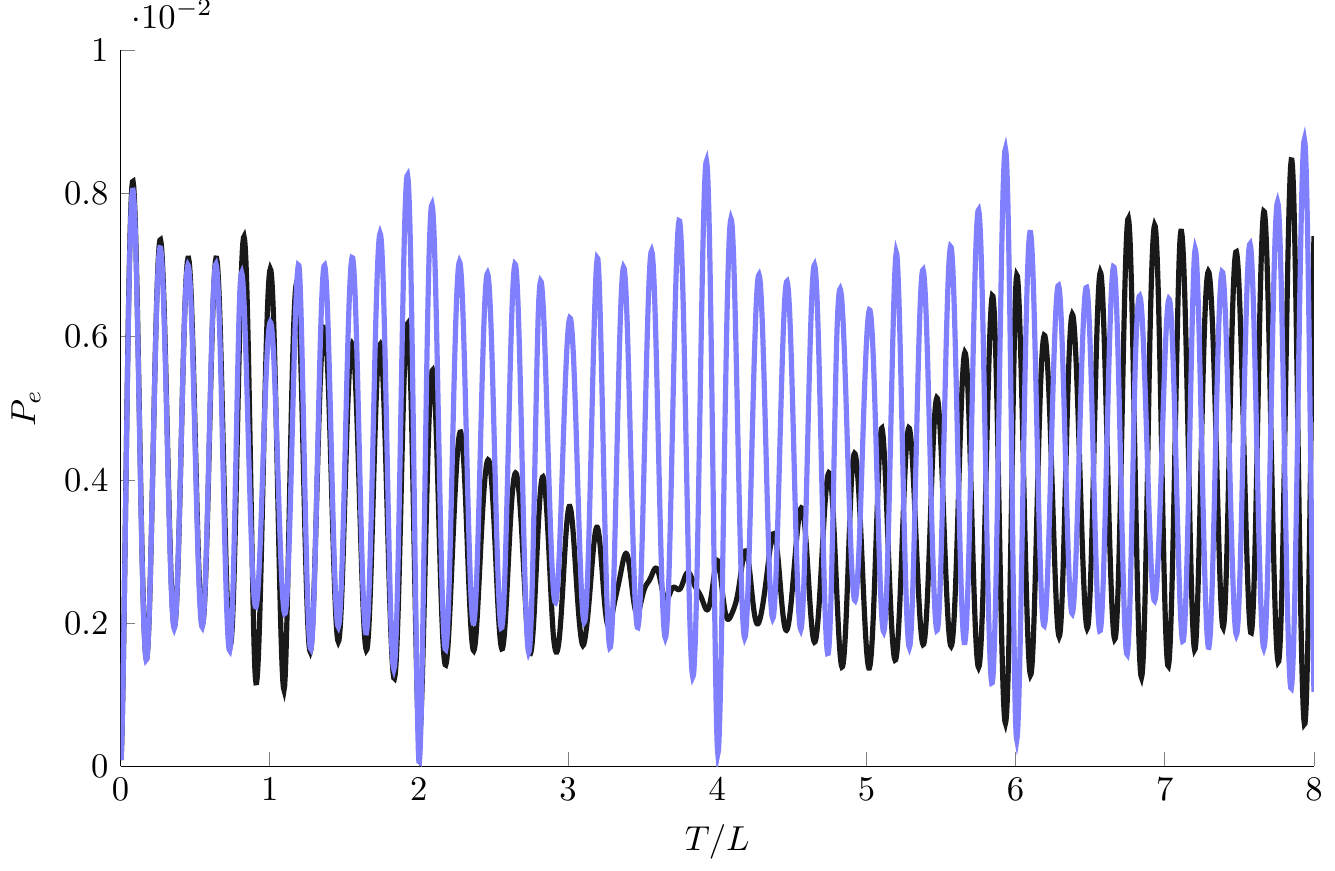}
  \caption{(Color online.) Position dependence of vacuum excitation probability in a Dirichlet cavity. A harmonic oscillator detector, resonant with the 10th field mode ($ \Omega= 10 \pi/L$) is initialized in its ground state and then coupled to the field for a time $t=0...T$. The plot shows the probability $P_e$ to find the detector in a state other than its ground state after the interaction with the field. For the black curve, whose oscillation amplitude diminishes towards 0.03 around $T=3.75$, the detector is located at $x=0.55L$ which is a maximum of the resonant 10th field mode. The light blue curve shows the detector located at $x=0.6L$ which is a node of the resonant field mode. The coupling constant is $\lambda=0.075$ For the numerical calculations $N=200$ field modes wereused.}
  \label{fig:app_vacuum_Pe}
\end{figure}

\begin{figure}
\centering
\includegraphics[width=\columnwidth]{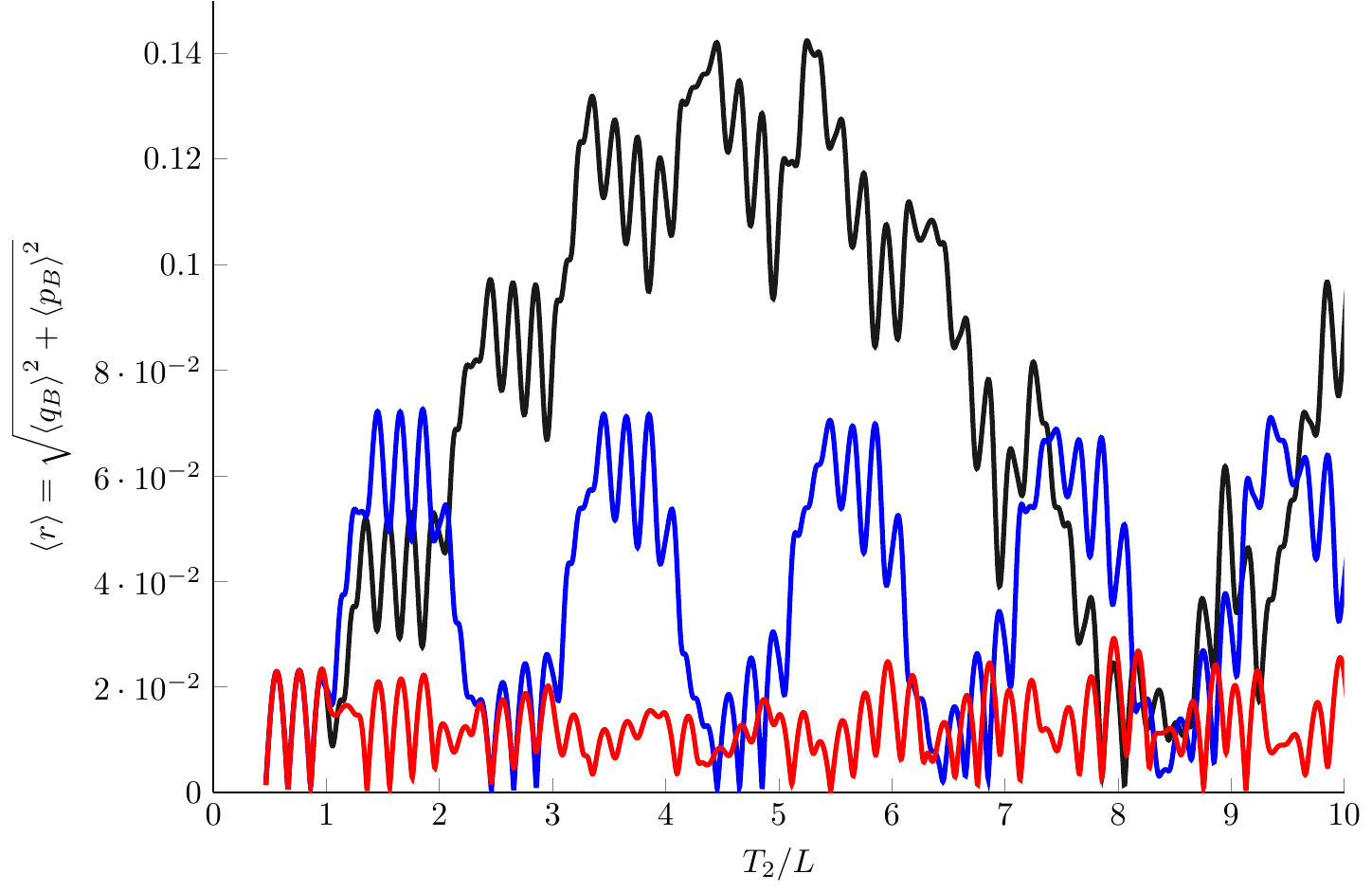}
  \caption{(Color online.) 
  Mean displacement $\exptval{r}$ 
  of final receiver state when the sender is initialized in a state with $\exptval{q_A}=0$ and $\exptval{p_A}=1$, i.e., with a non-zero displacement in canonical position. The plot shows different combinations of sender and receiver being located at maxima or node points of the field mode they are resonant with. The detectors'  energy spacing is $ \Omega=10 \pi/L$, i.e., they are resonant with the 10th cavity mode. The upper (black) curve corresponds to the sender being located at $a=0.45L$ and the receiver at $b=0.55L$ both of which are maxima of the 10th cavity mode. The middle (blue) curve correspond to both sender ($a=0.5L$) and receiver ($b=0.6L$) being located at node points. The lower (red) curve shows the sender at a node point ($a=0.5L$) and the receiver at a maximum ($b=0.55L$). The sender couples to the field for $t=0...0.3L$, the receiver for $t=0.46L...T_2$. The coupling constant is $\lambda=0.075$. For the numerical calculations $N=200$ field modes were used.
  }
  \label{fig:app_lin_mean_pA}
\end{figure}

\begin{figure}
\centering
\includegraphics[width=\columnwidth]{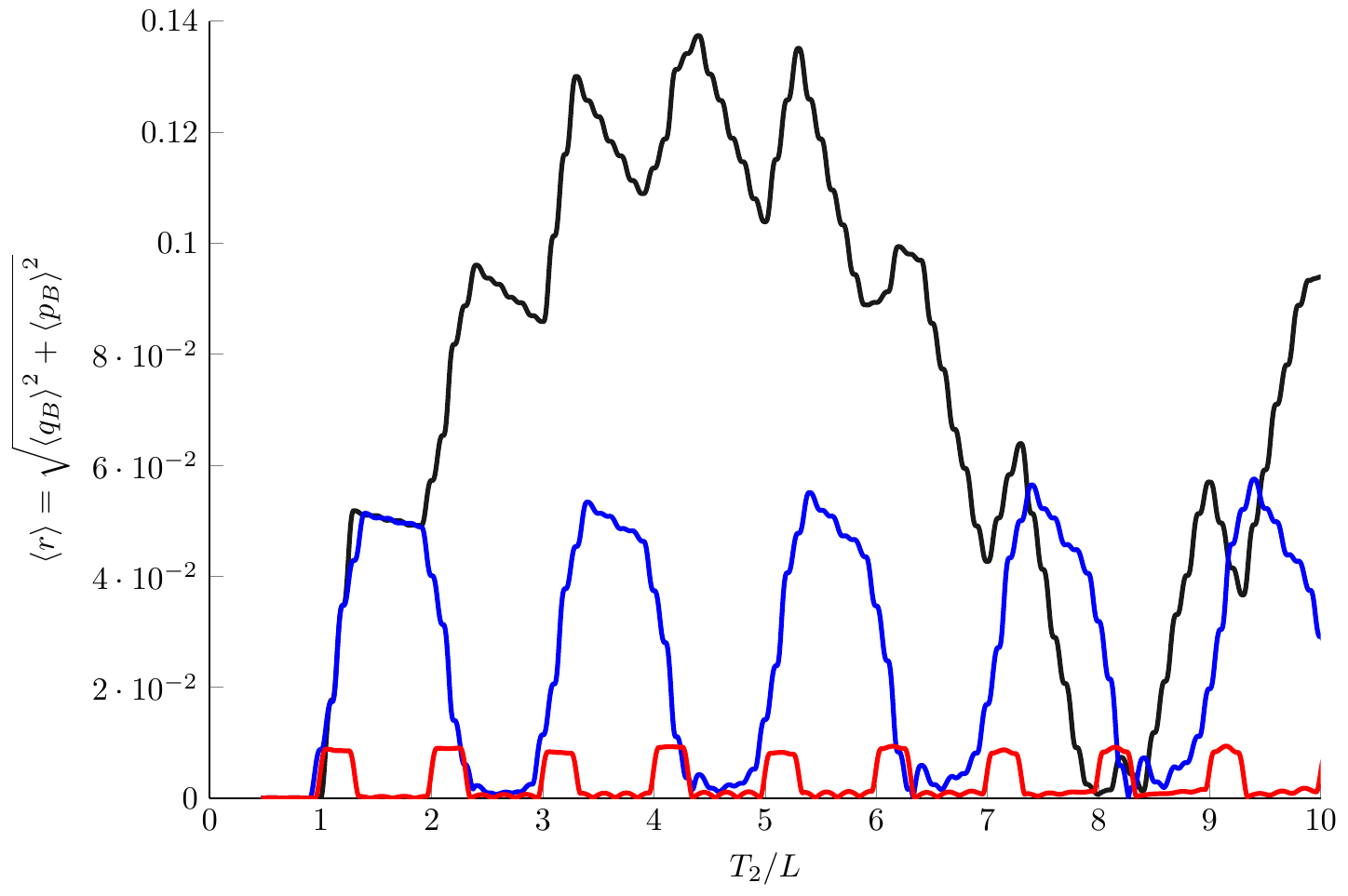}
  \caption{(Color online.)   Mean displacement $\exptval{r}$ 
  of final receiver state when the sender is initialized in a state with $\exptval{q_A}=1$ and $\exptval{p_A}=0$, i.e., with a non-zero displacement in canonical position. Otherwise setup identical to Fig.\@ \ref{fig:app_lin_mean_pA}.
  }
  \label{fig:app_lin_mean_xA}
\end{figure}

\begin{figure}
\centering
\includegraphics[width=\columnwidth]{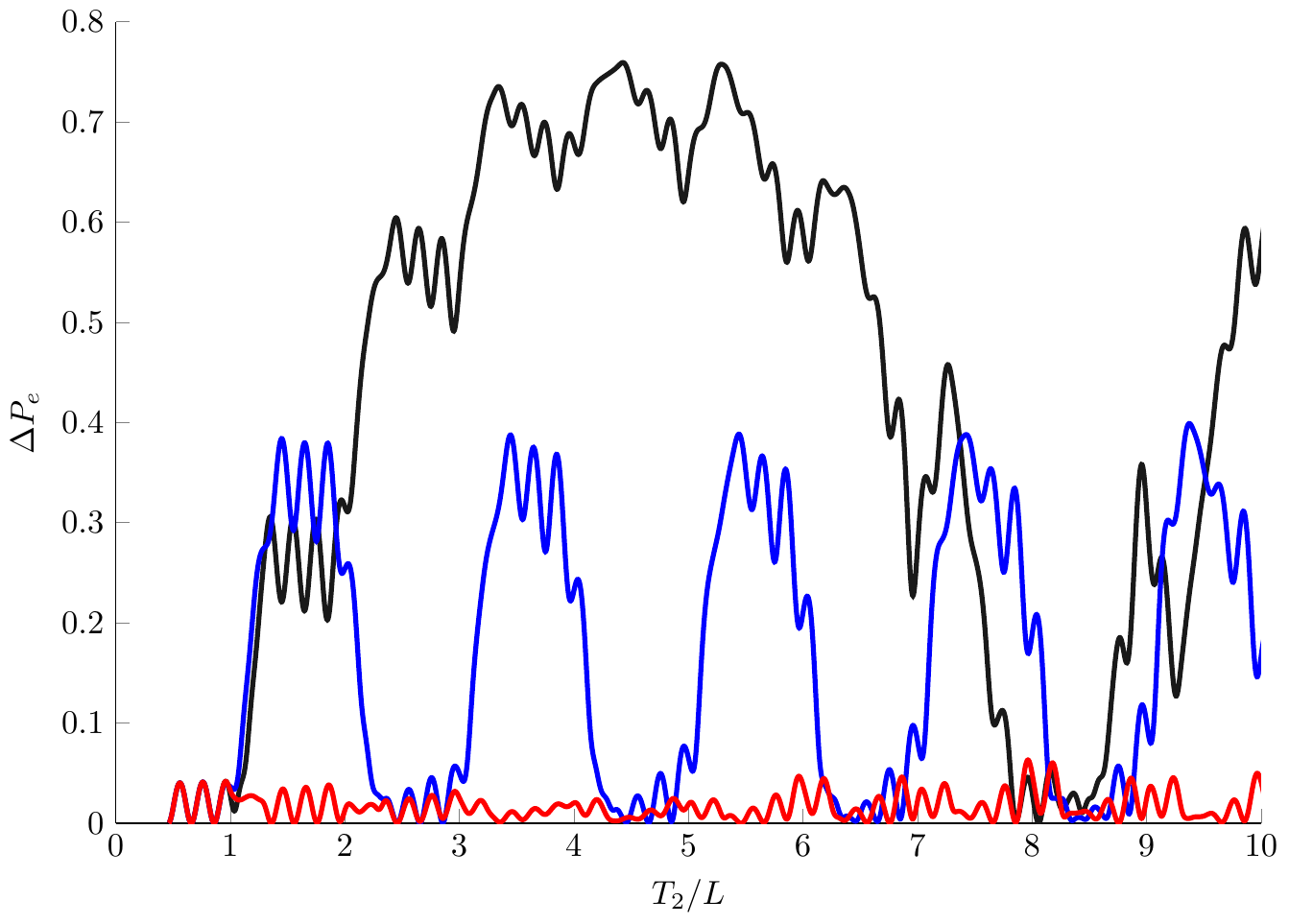}
  \caption{(Color online.) Signaling with zero-mean Gaussian states. The plot shows $ \Delta P_e= P_e^{\text{sig}} - P_e^{\text{vac}}$, which is the difference between the receiver's excitation probability  $P_e^{\text{sig}}$ in the signaling scenario and the receiver's vacuum excitation probability (see Fig.\@ \ref{fig:app_vacuum_Pe}). The sender was initialized in a thermal state with $ \Omega/T=6\cdot10^{-3}$. All other parameters and the detector locations are identical to Fig.\@ \ref{fig:app_lin_mean_pA}.
  }
  \label{fig:app_lin_thermal_long}
\end{figure}

The field inside a Dirichlet cavity is not translationally invariant since the individual fields modes exhibit nodes and maxima at different points in the cavity.  This also affects the coupling between a detector and the field. 
For example, the probability to excite a single detector by coupling it to the vacuum of the field for a while depends on the position of the detector inside the cavity, as shown in Fig.\@ \ref{fig:app_vacuum_Pe}. (See also \cite{jonsson_quantum_2014} for the excitation probability of a two-level detector.) 

Also the signaling between detectors is affected by the sender's and the receiver's position inside the cavity. Fig.\@ \ref{fig:app_lin_mean_pA} and Fig.\@ \ref{fig:app_lin_mean_xA} show how  the receiver's final mean displacement is affected, depending on whether sender and receiver are located at maxima or at node points of the cavity mode which their detector is resonant with. Fig.\@ \ref{fig:app_lin_mean_pA} is hereby extending the setup of Fig.\@ \ref{fig:timelike_mean_radial} to longer coupling times of the receiver's detector.

We see that the strongest displacement arises when both sender and receiver are located at  maxima of the resonant mode. When the sender and receiver both are at node points the signals are weaker, but still stronger than when when only the sender is located at a node point and the receiver is located at a maximum. 
(In the latter setup, even when reflected lightrays connect the sender to the receiver, the effects are not larger than for strictly timelike separation between sender and receiver.)

However, for strictly timelike separation between sender and receiver, the location of sender and receiver has no influence and there is no difference in the final displacement of the sender.

When signaling with zero-mean Gaussian states the effect of the detectors' position on the excitation probability of the receiver is very similar to the effect on the mean displacement. Fig.\@ \ref{fig:app_lin_thermal_long} shows  $ \Delta P_e= P_e^{\text{sig}} - P_e^{\text{vac}}$, which is the difference between the receiver's excitation probability  $P_e^{\text{sig}}$ in the signaling scenario and the receiver's excitation probability $P_e^{\text{vac}}$ when coupled only to the vacuum at the same location  as in Fig.\@ \ref{fig:app_vacuum_Pe}.

In contrast to the mean displacement $\exptval{r}$, the difference in excitation probability $ \Delta P_e$ shows some dependence on the exact location of sender and receiver in the cavity. However the size of this effect is negligible. For example, in the setup of Fig.\@ \ref{fig:app_lin_thermal_long} the differences in the value of $ \Delta P_e$ for different locations of sender and receiver are  $< 2\cdot10^{-5}$  whereas  $ \Delta P_e$  ranges up to $4\cdot 10^{-2}$.

\section{Timelike signaling between  detectors resonant with the fundamental cavity mode}\label{app:basemode}

\begin{figure}
\centering
\includegraphics[width=\columnwidth]{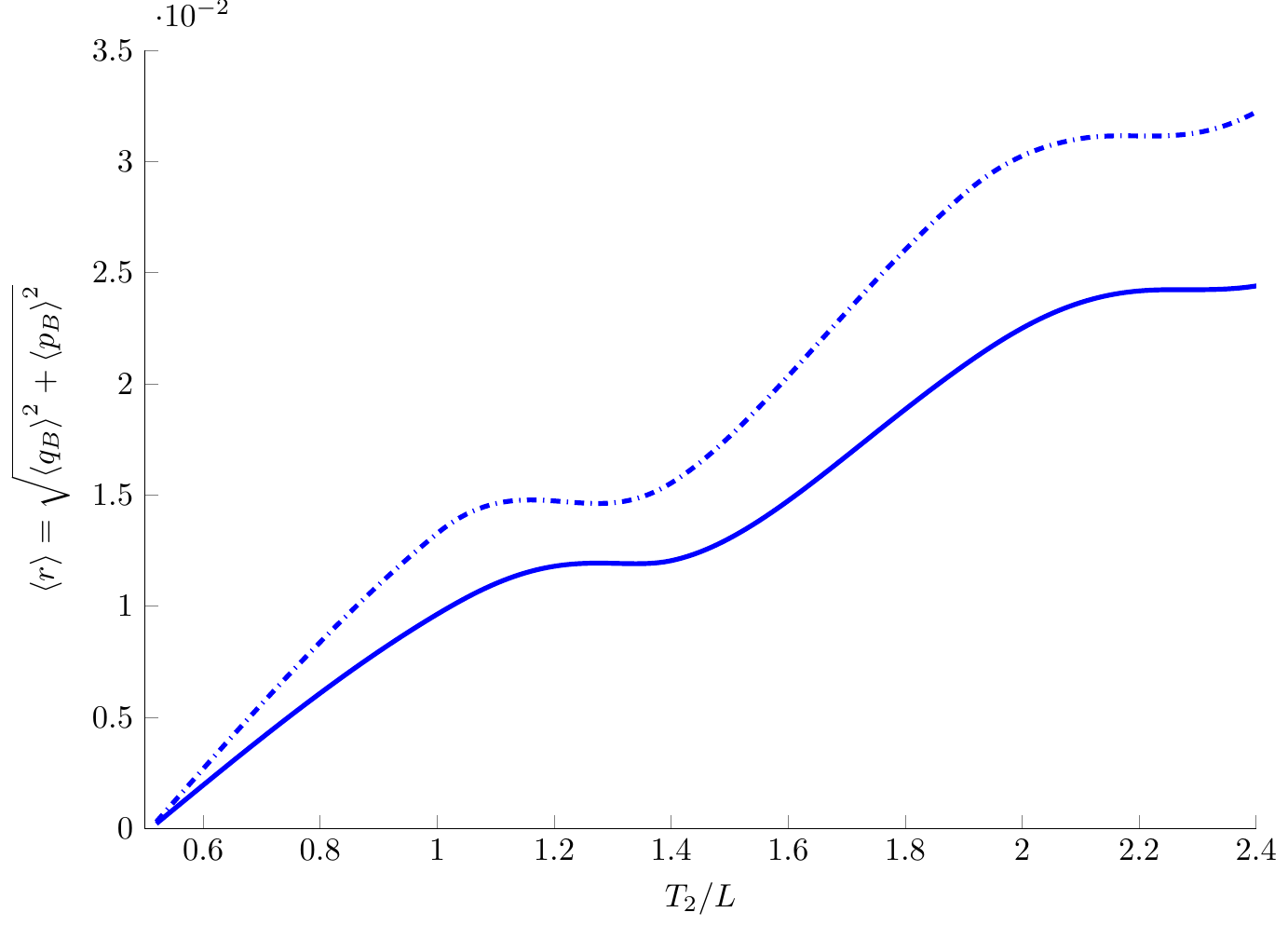}
  \caption{(Color online.) Mean displacement $\exptval{r}$ 
  of the receiver's final state, for detectors resonant with the fundamental cavity mode. The dotted line shows the displacement resulting from a initial sender state with $\exptval{q_A}=1$ and $\exptval{p_A}=0$, i.e., non-vanishing mean in canonical position. The other line shows an sender initial state with $\exptval{q_A}=0$ and $\exptval{p_A}=1$. The sender is located at $a=0.45L$, and coupled to the field for $t=0...0.4L$. The receiver is located at $b=0.55L$ and coupled to the field for $t=0.51L...T2$. Sender and receiver are timelike separated if $T_2<L$. Both detectors have $ \Omega= \pi/L$ and $ \lambda=0.01$. And $N=200$ cavity modes were used for the numerical calculations.
  }
  \label{fig:app_N_mean_radial}
\end{figure}

\begin{figure}
\centering
\includegraphics[width=\columnwidth]{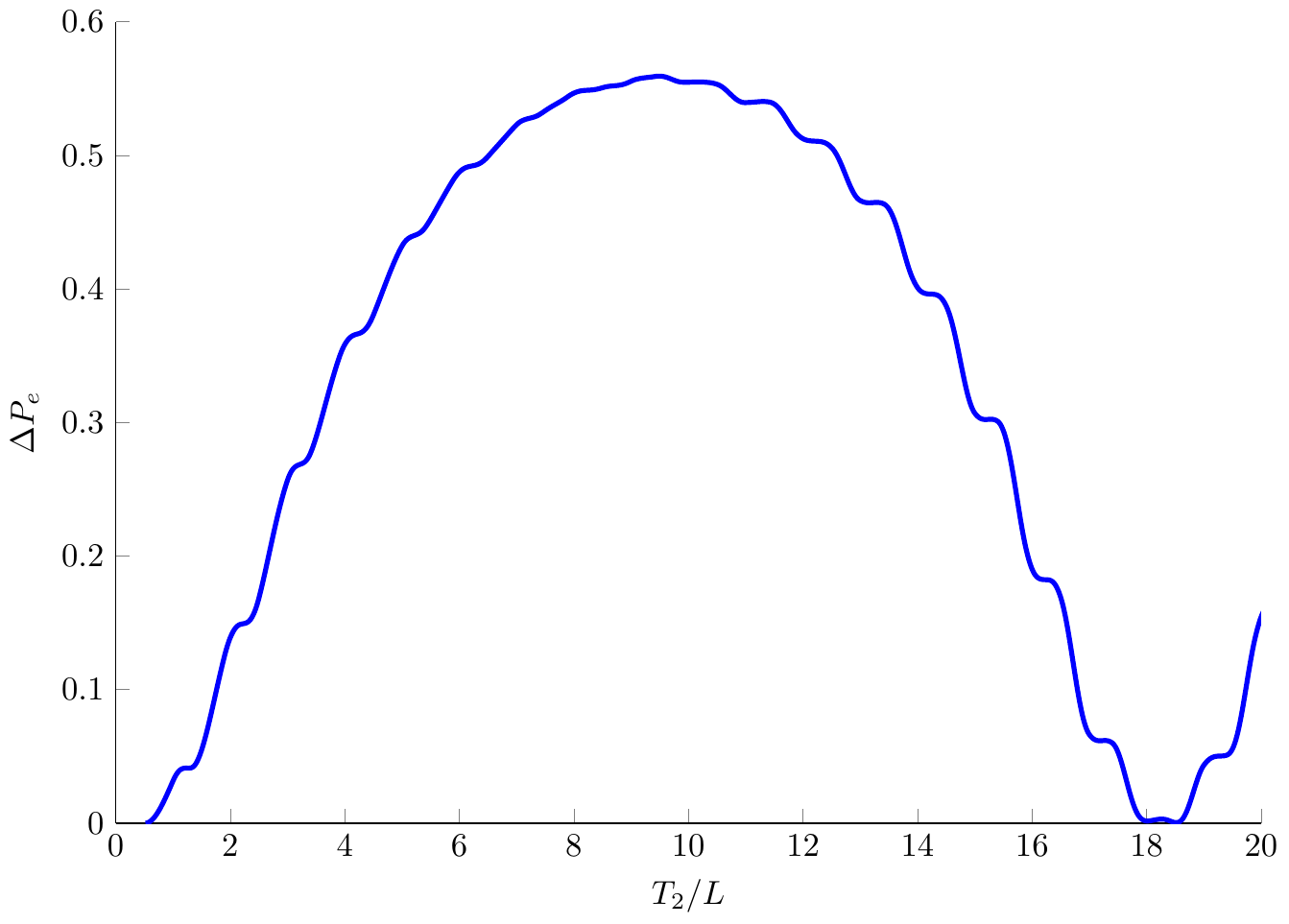}
  \caption{(Color online.) Signaling via zero-mean Gaussian states between detectors resonant with the fundamental cavity mode. The plot shows $ \Delta P_e= P_e^{\text{sig}} - P_e^{\text{vac}}$ the difference of excitation probability between the signaling scenario and coupling to the vacuum. (Compare Fig.\@ \ref{fig:app_lin_thermal_long}). The sender is initialized in a thermal state with $ \Omega/T=4\cdot 10^{-3}$. All other parameters are identical to Fig.\@ \ref{fig:app_N_mean_radial}. 
  Signaling arises already at timelike separations for  $T_2<L$.
  }
  \label{fig:app_N_thermal}
\end{figure}

In order to study timelike signaling via fields inside a cavity it appears natural to choose the cavity to be large, such that the receiver and sender can each  couple to the field for some time, before lightrays emanating from the sender and reflected by the cavity walls connect the sender to the receiver. If the interaction time of sender and receiver is to be on the order of a few detector periods $ \Omega/(2 \pi)$, then the detectors need to be resonant with higher modes of the cavity. For this reason we choose the detectors to be resonant with the 10th field mode in the numerical examples above.

It is still interesting to ask if timelike signaling also occurs between detectors that are resonant with the base mode of the cavity, i.e., have an energy level spacing of $ \Omega= \pi/L$. Fig.\@ \ref{fig:app_N_mean_radial} and Fig.\@ \ref{fig:app_N_thermal} answer this question in the affirmative.
They show the sender’s effect on the receiver’s final mean displacement, and on the receiver’s excitation probability for a zero-mean initial state of the sender. In this setup, in order to allow for timelike separation between sender and receiver, the sender is coupled to the field for less than half a detector period.

Fig.\@ \ref{fig:app_N_mean_radial} shows that unlike  detectors being resonant to higher field modes (Fig.\@ \ref{fig:timelike_mean_radial}), a displacement of the receiver's mean arises both from displacement in the sender's initial canonical momentum $\exptval{p_A}$, as well as in  position $\exptval{q_A}$. Whereas the mean displacement continues to grow for longer coupling times of the receiver, it already reaches the percentile range while sender and receiver are still strictly timelike separated.

The long time behaviour of the mean displacement is again similar to the behaviour of the excitation probability for zero-mean states which is given in Fig.\@ \ref{fig:app_N_thermal}. In Fig.\@ \ref{fig:app_N_thermal} we see the influence of a sender using a thermal, zero-mean Gaussian state on the excitation probability of the sender, analogous to Fig.\@ \ref{fig:app_lin_thermal_long}. Whereas $ \Delta P_e$ grows up to values of about $ \Delta P_e\approx 0.55$ for long coupling times, an influence on the order of $ \Delta P_e\approx 0.04$ is already visible for timelike separations between sender and receiver at the chosen parameter values.

\bibliography{energywhereinfocannotvpubl}

\begin{thebibliography}{54}%
\makeatletter
\providecommand \@ifxundefined [1]{%
 \@ifx{#1\undefined}
}%
\providecommand \@ifnum [1]{%
 \ifnum #1\expandafter \@firstoftwo
 \else \expandafter \@secondoftwo
 \fi
}%
\providecommand \@ifx [1]{%
 \ifx #1\expandafter \@firstoftwo
 \else \expandafter \@secondoftwo
 \fi
}%
\providecommand \natexlab [1]{#1}%
\providecommand \enquote  [1]{``#1''}%
\providecommand \bibnamefont  [1]{#1}%
\providecommand \bibfnamefont [1]{#1}%
\providecommand \citenamefont [1]{#1}%
\providecommand \href@noop [0]{\@secondoftwo}%
\providecommand \href [0]{\begingroup \@sanitize@url \@href}%
\providecommand \@href[1]{\@@startlink{#1}\@@href}%
\providecommand \@@href[1]{\endgroup#1\@@endlink}%
\providecommand \@sanitize@url [0]{\catcode `\\12\catcode `\$12\catcode
  `\&12\catcode `\#12\catcode `\^12\catcode `\_12\catcode `\%12\relax}%
\providecommand \@@startlink[1]{}%
\providecommand \@@endlink[0]{}%
\providecommand \url  [0]{\begingroup\@sanitize@url \@url }%
\providecommand \@url [1]{\endgroup\@href {#1}{\urlprefix }}%
\providecommand \urlprefix  [0]{URL }%
\providecommand \Eprint [0]{\href }%
\providecommand \doibase [0]{http://dx.doi.org/}%
\providecommand \selectlanguage [0]{\@gobble}%
\providecommand \bibinfo  [0]{\@secondoftwo}%
\providecommand \bibfield  [0]{\@secondoftwo}%
\providecommand \translation [1]{[#1]}%
\providecommand \BibitemOpen [0]{}%
\providecommand \bibitemStop [0]{}%
\providecommand \bibitemNoStop [0]{.\EOS\space}%
\providecommand \EOS [0]{\spacefactor3000\relax}%
\providecommand \BibitemShut  [1]{\csname bibitem#1\endcsname}%
\let\auto@bib@innerbib\@empty
\bibitem [{\citenamefont {Mathur}(2009)}]{mathur_information_2009}%
  \BibitemOpen
  \bibfield  {author} {\bibinfo {author} {\bibfnamefont {S.~D.}\ \bibnamefont
  {Mathur}},\ }\href {\doibase 10.1088/0264-9381/26/22/224001} {\bibfield
  {journal} {\bibinfo  {journal} {Classical and Quantum Gravity}\ }\textbf
  {\bibinfo {volume} {26}},\ \bibinfo {pages} {224001} (\bibinfo {year}
  {2009})},\ \bibinfo {note} {arXiv: 0909.1038}\BibitemShut {NoStop}%
\bibitem [{\citenamefont {Braunstein}\ \emph {et~al.}(2013)\citenamefont
  {Braunstein}, \citenamefont {Pirandola},\ and\ \citenamefont
  {Życzkowski}}]{braunstein_better_2013}%
  \BibitemOpen
  \bibfield  {author} {\bibinfo {author} {\bibfnamefont {S.~L.}\ \bibnamefont
  {Braunstein}}, \bibinfo {author} {\bibfnamefont {S.}~\bibnamefont
  {Pirandola}}, \ and\ \bibinfo {author} {\bibfnamefont {K.}~\bibnamefont
  {Życzkowski}},\ }\href {\doibase 10.1103/PhysRevLett.110.101301} {\bibfield
  {journal} {\bibinfo  {journal} {Physical Review Letters}\ }\textbf {\bibinfo
  {volume} {110}},\ \bibinfo {pages} {101301} (\bibinfo {year}
  {2013})}\BibitemShut {NoStop}%
\bibitem [{\citenamefont {Almheiri}\ \emph {et~al.}(2013)\citenamefont
  {Almheiri}, \citenamefont {Marolf}, \citenamefont {Polchinski},\ and\
  \citenamefont {Sully}}]{almheiri_black_2013}%
  \BibitemOpen
  \bibfield  {author} {\bibinfo {author} {\bibfnamefont {A.}~\bibnamefont
  {Almheiri}}, \bibinfo {author} {\bibfnamefont {D.}~\bibnamefont {Marolf}},
  \bibinfo {author} {\bibfnamefont {J.}~\bibnamefont {Polchinski}}, \ and\
  \bibinfo {author} {\bibfnamefont {J.}~\bibnamefont {Sully}},\ }\href
  {\doibase 10.1007/JHEP02(2013)062} {\bibfield  {journal} {\bibinfo  {journal}
  {Journal of High Energy Physics}\ }\textbf {\bibinfo {volume} {2013}}
  (\bibinfo {year} {2013}),\ 10.1007/JHEP02(2013)062},\ \bibinfo {note} {arXiv:
  1207.3123}\BibitemShut {NoStop}%
\bibitem [{\citenamefont {Bradler}\ \emph {et~al.}(2012)\citenamefont
  {Bradler}, \citenamefont {Hayden},\ and\ \citenamefont
  {Panangaden}}]{bradler_quantum_2012}%
  \BibitemOpen
  \bibfield  {author} {\bibinfo {author} {\bibfnamefont {K.}~\bibnamefont
  {Bradler}}, \bibinfo {author} {\bibfnamefont {P.}~\bibnamefont {Hayden}}, \
  and\ \bibinfo {author} {\bibfnamefont {P.}~\bibnamefont {Panangaden}},\
  }\href {\doibase 10.1007/s00220-012-1476-1} {\bibfield  {journal} {\bibinfo
  {journal} {Communications in Mathematical Physics}\ }\textbf {\bibinfo
  {volume} {312}},\ \bibinfo {pages} {361} (\bibinfo {year} {2012})},\ \bibinfo
  {note} {arXiv: 1007.0997}\BibitemShut {NoStop}%
\bibitem [{\citenamefont {Cliche}\ and\ \citenamefont
  {Kempf}(2010)}]{cliche_relativistic_2010}%
  \BibitemOpen
  \bibfield  {author} {\bibinfo {author} {\bibfnamefont {M.}~\bibnamefont
  {Cliche}}\ and\ \bibinfo {author} {\bibfnamefont {A.}~\bibnamefont {Kempf}},\
  }\href {\doibase 10.1103/PhysRevA.81.012330} {\bibfield  {journal} {\bibinfo
  {journal} {Physical Review A}\ }\textbf {\bibinfo {volume} {81}},\ \bibinfo
  {pages} {012330} (\bibinfo {year} {2010})}\BibitemShut {NoStop}%
\bibitem [{\citenamefont {Downes}\ \emph {et~al.}(2013)\citenamefont {Downes},
  \citenamefont {Ralph},\ and\ \citenamefont {Walk}}]{downes_quantum_2013}%
  \BibitemOpen
  \bibfield  {author} {\bibinfo {author} {\bibfnamefont {T.~G.}\ \bibnamefont
  {Downes}}, \bibinfo {author} {\bibfnamefont {T.~C.}\ \bibnamefont {Ralph}}, \
  and\ \bibinfo {author} {\bibfnamefont {N.}~\bibnamefont {Walk}},\ }\href
  {\doibase 10.1103/PhysRevA.87.012327} {\bibfield  {journal} {\bibinfo
  {journal} {Physical Review A}\ }\textbf {\bibinfo {volume} {87}},\ \bibinfo
  {pages} {012327} (\bibinfo {year} {2013})}\BibitemShut {NoStop}%
\bibitem [{\citenamefont {Wilson}\ \emph {et~al.}(2011)\citenamefont {Wilson},
  \citenamefont {Johansson}, \citenamefont {Pourkabirian}, \citenamefont
  {Simoen}, \citenamefont {Johansson}, \citenamefont {Duty}, \citenamefont
  {Nori},\ and\ \citenamefont {Delsing}}]{wilson_observation_2011}%
  \BibitemOpen
  \bibfield  {author} {\bibinfo {author} {\bibfnamefont {C.~M.}\ \bibnamefont
  {Wilson}}, \bibinfo {author} {\bibfnamefont {G.}~\bibnamefont {Johansson}},
  \bibinfo {author} {\bibfnamefont {A.}~\bibnamefont {Pourkabirian}}, \bibinfo
  {author} {\bibfnamefont {M.}~\bibnamefont {Simoen}}, \bibinfo {author}
  {\bibfnamefont {J.~R.}\ \bibnamefont {Johansson}}, \bibinfo {author}
  {\bibfnamefont {T.}~\bibnamefont {Duty}}, \bibinfo {author} {\bibfnamefont
  {F.}~\bibnamefont {Nori}}, \ and\ \bibinfo {author} {\bibfnamefont
  {P.}~\bibnamefont {Delsing}},\ }\href {\doibase 10.1038/nature10561}
  {\bibfield  {journal} {\bibinfo  {journal} {Nature}\ }\textbf {\bibinfo
  {volume} {479}},\ \bibinfo {pages} {376} (\bibinfo {year}
  {2011})}\BibitemShut {NoStop}%
\bibitem [{\citenamefont {Friis}\ \emph {et~al.}(2013)\citenamefont {Friis},
  \citenamefont {Lee}, \citenamefont {Truong}, \citenamefont {Sabín},
  \citenamefont {Solano}, \citenamefont {Johansson},\ and\ \citenamefont
  {Fuentes}}]{friis_relativistic_2013}%
  \BibitemOpen
  \bibfield  {author} {\bibinfo {author} {\bibfnamefont {N.}~\bibnamefont
  {Friis}}, \bibinfo {author} {\bibfnamefont {A.~R.}\ \bibnamefont {Lee}},
  \bibinfo {author} {\bibfnamefont {K.}~\bibnamefont {Truong}}, \bibinfo
  {author} {\bibfnamefont {C.}~\bibnamefont {Sabín}}, \bibinfo {author}
  {\bibfnamefont {E.}~\bibnamefont {Solano}}, \bibinfo {author} {\bibfnamefont
  {G.}~\bibnamefont {Johansson}}, \ and\ \bibinfo {author} {\bibfnamefont
  {I.}~\bibnamefont {Fuentes}},\ }\href {\doibase
  10.1103/PhysRevLett.110.113602} {\bibfield  {journal} {\bibinfo  {journal}
  {Physical Review Letters}\ }\textbf {\bibinfo {volume} {110}},\ \bibinfo
  {pages} {113602} (\bibinfo {year} {2013})}\BibitemShut {NoStop}%
\bibitem [{\citenamefont {Lindkvist}\ \emph {et~al.}(2014)\citenamefont
  {Lindkvist}, \citenamefont {Sabín}, \citenamefont {Fuentes}, \citenamefont
  {Dragan}, \citenamefont {Svensson}, \citenamefont {Delsing},\ and\
  \citenamefont {Johansson}}]{lindkvist_twin_2014}%
  \BibitemOpen
  \bibfield  {author} {\bibinfo {author} {\bibfnamefont {J.}~\bibnamefont
  {Lindkvist}}, \bibinfo {author} {\bibfnamefont {C.}~\bibnamefont {Sabín}},
  \bibinfo {author} {\bibfnamefont {I.}~\bibnamefont {Fuentes}}, \bibinfo
  {author} {\bibfnamefont {A.}~\bibnamefont {Dragan}}, \bibinfo {author}
  {\bibfnamefont {I.-M.}\ \bibnamefont {Svensson}}, \bibinfo {author}
  {\bibfnamefont {P.}~\bibnamefont {Delsing}}, \ and\ \bibinfo {author}
  {\bibfnamefont {G.}~\bibnamefont {Johansson}},\ }\href {\doibase
  10.1103/PhysRevA.90.052113} {\bibfield  {journal} {\bibinfo  {journal}
  {Physical Review A}\ }\textbf {\bibinfo {volume} {90}} (\bibinfo {year}
  {2014}),\ 10.1103/PhysRevA.90.052113},\ \bibinfo {note} {arXiv:
  1401.0129}\BibitemShut {NoStop}%
\bibitem [{\citenamefont {Reznik}\ \emph {et~al.}(2005)\citenamefont {Reznik},
  \citenamefont {Retzker},\ and\ \citenamefont
  {Silman}}]{reznik_violating_2005}%
  \BibitemOpen
  \bibfield  {author} {\bibinfo {author} {\bibfnamefont {B.}~\bibnamefont
  {Reznik}}, \bibinfo {author} {\bibfnamefont {A.}~\bibnamefont {Retzker}}, \
  and\ \bibinfo {author} {\bibfnamefont {J.}~\bibnamefont {Silman}},\ }\href
  {\doibase 10.1103/PhysRevA.71.042104} {\bibfield  {journal} {\bibinfo
  {journal} {Physical Review A}\ }\textbf {\bibinfo {volume} {71}},\ \bibinfo
  {pages} {042104} (\bibinfo {year} {2005})}\BibitemShut {NoStop}%
\bibitem [{\citenamefont {Martín-Martínez}\ and\ \citenamefont
  {Menicucci}(2014)}]{martin-martinez_entanglement_2014}%
  \BibitemOpen
  \bibfield  {author} {\bibinfo {author} {\bibfnamefont {E.}~\bibnamefont
  {Martín-Martínez}}\ and\ \bibinfo {author} {\bibfnamefont {N.~C.}\
  \bibnamefont {Menicucci}},\ }\href {\doibase 10.1088/0264-9381/31/21/214001}
  {\bibfield  {journal} {\bibinfo  {journal} {Classical and Quantum Gravity}\
  }\textbf {\bibinfo {volume} {31}},\ \bibinfo {pages} {214001} (\bibinfo
  {year} {2014})}\BibitemShut {NoStop}%
\bibitem [{\citenamefont {Steeg}\ and\ \citenamefont
  {Menicucci}(2009)}]{steeg_entangling_2009}%
  \BibitemOpen
  \bibfield  {author} {\bibinfo {author} {\bibfnamefont {G.~V.}\ \bibnamefont
  {Steeg}}\ and\ \bibinfo {author} {\bibfnamefont {N.~C.}\ \bibnamefont
  {Menicucci}},\ }\href {\doibase 10.1103/PhysRevD.79.044027} {\bibfield
  {journal} {\bibinfo  {journal} {Physical Review D}\ }\textbf {\bibinfo
  {volume} {79}},\ \bibinfo {pages} {044027} (\bibinfo {year}
  {2009})}\BibitemShut {NoStop}%
\bibitem [{\citenamefont {Cliche}\ and\ \citenamefont
  {Kempf}(2011)}]{cliche_vacuum_2011}%
  \BibitemOpen
  \bibfield  {author} {\bibinfo {author} {\bibfnamefont {M.}~\bibnamefont
  {Cliche}}\ and\ \bibinfo {author} {\bibfnamefont {A.}~\bibnamefont {Kempf}},\
  }\href {\doibase 10.1103/PhysRevD.83.045019} {\bibfield  {journal} {\bibinfo
  {journal} {Physical Review D}\ }\textbf {\bibinfo {volume} {83}},\ \bibinfo
  {pages} {045019} (\bibinfo {year} {2011})}\BibitemShut {NoStop}%
\bibitem [{\citenamefont {Salton}\ \emph {et~al.}(2015)\citenamefont {Salton},
  \citenamefont {Mann},\ and\ \citenamefont
  {Menicucci}}]{salton_acceleration-assisted_2015}%
  \BibitemOpen
  \bibfield  {author} {\bibinfo {author} {\bibfnamefont {G.}~\bibnamefont
  {Salton}}, \bibinfo {author} {\bibfnamefont {R.~B.}\ \bibnamefont {Mann}}, \
  and\ \bibinfo {author} {\bibfnamefont {N.~C.}\ \bibnamefont {Menicucci}},\
  }\href {\doibase 10.1088/1367-2630/17/3/035001} {\bibfield  {journal}
  {\bibinfo  {journal} {New Journal of Physics}\ }\textbf {\bibinfo {volume}
  {17}},\ \bibinfo {pages} {035001} (\bibinfo {year} {2015})}\BibitemShut
  {NoStop}%
\bibitem [{\citenamefont {Brown}(2013)}]{brown_thermal_2013}%
  \BibitemOpen
  \bibfield  {author} {\bibinfo {author} {\bibfnamefont {E.~G.}\ \bibnamefont
  {Brown}},\ }\href {\doibase 10.1103/PhysRevA.88.062336} {\bibfield  {journal}
  {\bibinfo  {journal} {Physical Review A}\ }\textbf {\bibinfo {volume} {88}},\
  \bibinfo {pages} {062336} (\bibinfo {year} {2013})}\BibitemShut {NoStop}%
\bibitem [{\citenamefont {Martín-Martínez}\ \emph
  {et~al.}(2013{\natexlab{a}})\citenamefont {Martín-Martínez}, \citenamefont
  {Brown}, \citenamefont {Donnelly},\ and\ \citenamefont
  {Kempf}}]{martin-martinez_sustainable_2013}%
  \BibitemOpen
  \bibfield  {author} {\bibinfo {author} {\bibfnamefont {E.}~\bibnamefont
  {Martín-Martínez}}, \bibinfo {author} {\bibfnamefont {E.~G.}\ \bibnamefont
  {Brown}}, \bibinfo {author} {\bibfnamefont {W.}~\bibnamefont {Donnelly}}, \
  and\ \bibinfo {author} {\bibfnamefont {A.}~\bibnamefont {Kempf}},\ }\href
  {\doibase 10.1103/PhysRevA.88.052310} {\bibfield  {journal} {\bibinfo
  {journal} {Physical Review A}\ }\textbf {\bibinfo {volume} {88}},\ \bibinfo
  {pages} {052310} (\bibinfo {year} {2013}{\natexlab{a}})}\BibitemShut
  {NoStop}%
\bibitem [{\citenamefont {Sabín}\ \emph {et~al.}(2012)\citenamefont {Sabín},
  \citenamefont {Peropadre}, \citenamefont {del Rey},\ and\ \citenamefont
  {Martín-Martínez}}]{sabin_extracting_2012}%
  \BibitemOpen
  \bibfield  {author} {\bibinfo {author} {\bibfnamefont {C.}~\bibnamefont
  {Sabín}}, \bibinfo {author} {\bibfnamefont {B.}~\bibnamefont {Peropadre}},
  \bibinfo {author} {\bibfnamefont {M.}~\bibnamefont {del Rey}}, \ and\
  \bibinfo {author} {\bibfnamefont {E.}~\bibnamefont {Martín-Martínez}},\
  }\href {\doibase 10.1103/PhysRevLett.109.033602} {\bibfield  {journal}
  {\bibinfo  {journal} {Physical Review Letters}\ }\textbf {\bibinfo {volume}
  {109}},\ \bibinfo {pages} {033602} (\bibinfo {year} {2012})}\BibitemShut
  {NoStop}%
\bibitem [{\citenamefont {Hotta}\ \emph {et~al.}(2014)\citenamefont {Hotta},
  \citenamefont {Matsumoto},\ and\ \citenamefont {Yusa}}]{hotta_quantum_2014}%
  \BibitemOpen
  \bibfield  {author} {\bibinfo {author} {\bibfnamefont {M.}~\bibnamefont
  {Hotta}}, \bibinfo {author} {\bibfnamefont {J.}~\bibnamefont {Matsumoto}}, \
  and\ \bibinfo {author} {\bibfnamefont {G.}~\bibnamefont {Yusa}},\ }\href
  {\doibase 10.1103/PhysRevA.89.012311} {\bibfield  {journal} {\bibinfo
  {journal} {Physical Review A}\ }\textbf {\bibinfo {volume} {89}},\ \bibinfo
  {pages} {012311} (\bibinfo {year} {2014})}\BibitemShut {NoStop}%
\bibitem [{\citenamefont {Hotta}(2008)}]{hotta_quantum_2008}%
  \BibitemOpen
  \bibfield  {author} {\bibinfo {author} {\bibfnamefont {M.}~\bibnamefont
  {Hotta}},\ }\href {\doibase 10.1103/PhysRevD.78.045006} {\bibfield  {journal}
  {\bibinfo  {journal} {Physical Review D}\ }\textbf {\bibinfo {volume} {78}},\
  \bibinfo {pages} {045006} (\bibinfo {year} {2008})}\BibitemShut {NoStop}%
\bibitem [{\citenamefont {Hotta}(2010)}]{hotta_controlled_2010}%
  \BibitemOpen
  \bibfield  {author} {\bibinfo {author} {\bibfnamefont {M.}~\bibnamefont
  {Hotta}},\ }\href {\doibase 10.1103/PhysRevD.81.044025} {\bibfield  {journal}
  {\bibinfo  {journal} {Physical Review D}\ }\textbf {\bibinfo {volume} {81}},\
  \bibinfo {pages} {044025} (\bibinfo {year} {2010})}\BibitemShut {NoStop}%
\bibitem [{\citenamefont {Verdon-Akzam}\ \emph {et~al.}(2016)\citenamefont
  {Verdon-Akzam}, \citenamefont {Martín-Martínez},\ and\ \citenamefont
  {Kempf}}]{verdon-akzam_asymptotically_2016}%
  \BibitemOpen
  \bibfield  {author} {\bibinfo {author} {\bibfnamefont {G.}~\bibnamefont
  {Verdon-Akzam}}, \bibinfo {author} {\bibfnamefont {E.}~\bibnamefont
  {Martín-Martínez}}, \ and\ \bibinfo {author} {\bibfnamefont
  {A.}~\bibnamefont {Kempf}},\ }\href {\doibase 10.1103/PhysRevA.93.022308}
  {\bibfield  {journal} {\bibinfo  {journal} {Physical Review A}\ }\textbf
  {\bibinfo {volume} {93}},\ \bibinfo {pages} {022308} (\bibinfo {year}
  {2016})}\BibitemShut {NoStop}%
\bibitem [{\citenamefont {Jonsson}\ \emph {et~al.}(2015)\citenamefont
  {Jonsson}, \citenamefont {Martín-Martínez},\ and\ \citenamefont
  {Kempf}}]{jonsson_information_2015}%
  \BibitemOpen
  \bibfield  {author} {\bibinfo {author} {\bibfnamefont {R.~H.}\ \bibnamefont
  {Jonsson}}, \bibinfo {author} {\bibfnamefont {E.}~\bibnamefont
  {Martín-Martínez}}, \ and\ \bibinfo {author} {\bibfnamefont
  {A.}~\bibnamefont {Kempf}},\ }\href {\doibase 10.1103/PhysRevLett.114.110505}
  {\bibfield  {journal} {\bibinfo  {journal} {Physical Review Letters}\
  }\textbf {\bibinfo {volume} {114}},\ \bibinfo {pages} {110505} (\bibinfo
  {year} {2015})}\BibitemShut {NoStop}%
\bibitem [{\citenamefont {McLenaghan}(1974)}]{mclenaghan_validity_1974}%
  \BibitemOpen
  \bibfield  {author} {\bibinfo {author} {\bibfnamefont {R.~G.}\ \bibnamefont
  {McLenaghan}},\ }in\ \href
  {http://archive.numdam.org/article/AIHPA_1974__20_2_153_0.pdf} {\emph
  {\bibinfo {booktitle} {Annales de l'{IHP} {Physique} théorique}}},\
  Vol.~\bibinfo {volume} {20}\ (\bibinfo {year} {1974})\ pp.\ \bibinfo {pages}
  {153--188}\BibitemShut {NoStop}%
\bibitem [{\citenamefont {Czapor}\ and\ \citenamefont
  {McLenaghan}(2007)}]{czapor_hadamards_2007}%
  \BibitemOpen
  \bibfield  {author} {\bibinfo {author} {\bibfnamefont {S.~R.}\ \bibnamefont
  {Czapor}}\ and\ \bibinfo {author} {\bibfnamefont {R.~G.}\ \bibnamefont
  {McLenaghan}},\ }\href
  {http://www.actaphys.uj.edu.pl/_old/sup1/pdf/s1p0055.pdf} {\bibfield
  {journal} {\bibinfo  {journal} {ACTA PHYSICA POLONICA SERIES B}\ }\textbf
  {\bibinfo {volume} {1}},\ \bibinfo {pages} {55} (\bibinfo {year}
  {2007})}\BibitemShut {NoStop}%
\bibitem [{\citenamefont {Poisson}\ \emph {et~al.}(2011)\citenamefont
  {Poisson}, \citenamefont {Pound},\ and\ \citenamefont
  {Vega}}]{poisson_motion_2011}%
  \BibitemOpen
  \bibfield  {author} {\bibinfo {author} {\bibfnamefont {E.}~\bibnamefont
  {Poisson}}, \bibinfo {author} {\bibfnamefont {A.}~\bibnamefont {Pound}}, \
  and\ \bibinfo {author} {\bibfnamefont {I.}~\bibnamefont {Vega}},\ }\href
  {\doibase 10.12942/lrr-2011-7} {\bibfield  {journal} {\bibinfo  {journal}
  {Living Reviews in Relativity}\ }\textbf {\bibinfo {volume} {14}} (\bibinfo
  {year} {2011}),\ 10.12942/lrr-2011-7}\BibitemShut {NoStop}%
\bibitem [{\citenamefont {Blasco}\ \emph {et~al.}(2015)\citenamefont {Blasco},
  \citenamefont {Garay}, \citenamefont {Martín-Benito},\ and\ \citenamefont
  {Martín-Martínez}}]{blasco_violation_2015}%
  \BibitemOpen
  \bibfield  {author} {\bibinfo {author} {\bibfnamefont {A.}~\bibnamefont
  {Blasco}}, \bibinfo {author} {\bibfnamefont {L.~J.}\ \bibnamefont {Garay}},
  \bibinfo {author} {\bibfnamefont {M.}~\bibnamefont {Martín-Benito}}, \ and\
  \bibinfo {author} {\bibfnamefont {E.}~\bibnamefont {Martín-Martínez}},\
  }\href {\doibase 10.1103/PhysRevLett.114.141103} {\bibfield  {journal}
  {\bibinfo  {journal} {Physical Review Letters}\ }\textbf {\bibinfo {volume}
  {114}},\ \bibinfo {pages} {141103} (\bibinfo {year} {2015})}\BibitemShut
  {NoStop}%
\bibitem [{\citenamefont {Blasco}\ \emph {et~al.}(2016)\citenamefont {Blasco},
  \citenamefont {Garay}, \citenamefont {Martín-Benito},\ and\ \citenamefont
  {Martín-Martínez}}]{blasco_timelike_2016}%
  \BibitemOpen
  \bibfield  {author} {\bibinfo {author} {\bibfnamefont {A.}~\bibnamefont
  {Blasco}}, \bibinfo {author} {\bibfnamefont {L.~J.}\ \bibnamefont {Garay}},
  \bibinfo {author} {\bibfnamefont {M.}~\bibnamefont {Martín-Benito}}, \ and\
  \bibinfo {author} {\bibfnamefont {E.}~\bibnamefont {Martín-Martínez}},\
  }\href {\doibase 10.1103/PhysRevD.93.024055} {\bibfield  {journal} {\bibinfo
  {journal} {Physical Review D}\ }\textbf {\bibinfo {volume} {93}},\ \bibinfo
  {pages} {024055} (\bibinfo {year} {2016})},\ \bibinfo {note} {bibtex:
  blasco\_timelike\_2016}\BibitemShut {NoStop}%
\bibitem [{\citenamefont {Unruh}(1976)}]{unruh_notes_1976}%
  \BibitemOpen
  \bibfield  {author} {\bibinfo {author} {\bibfnamefont {W.~G.}\ \bibnamefont
  {Unruh}},\ }\href {\doibase 10.1103/PhysRevD.14.870} {\bibfield  {journal}
  {\bibinfo  {journal} {Physical Review D}\ }\textbf {\bibinfo {volume} {14}},\
  \bibinfo {pages} {870} (\bibinfo {year} {1976})}\BibitemShut {NoStop}%
\bibitem [{\citenamefont {DeWitt}(1979)}]{hawking_quantum_1979}%
  \BibitemOpen
  \bibfield  {author} {\bibinfo {author} {\bibfnamefont {B.~S.}\ \bibnamefont
  {DeWitt}},\ }in\ \href@noop {} {\emph {\bibinfo {booktitle} {General
  relativity : an {Einstein} centenary survey}}},\ \bibinfo {editor} {edited
  by\ \bibinfo {editor} {\bibfnamefont {S.}~\bibnamefont {Hawking}}\ and\
  \bibinfo {editor} {\bibfnamefont {W.}~\bibnamefont {Israel}}}\ (\bibinfo
  {publisher} {Cambridge University Press},\ \bibinfo {address} {Cambridge Eng;
  New York},\ \bibinfo {year} {1979})\ p.\ \bibinfo {pages} {680}\BibitemShut
  {NoStop}%
\bibitem [{\citenamefont {Martín-Martínez}\ \emph
  {et~al.}(2013{\natexlab{b}})\citenamefont {Martín-Martínez}, \citenamefont
  {Montero},\ and\ \citenamefont {del Rey}}]{martin-martinez_wavepacket_2013}%
  \BibitemOpen
  \bibfield  {author} {\bibinfo {author} {\bibfnamefont {E.}~\bibnamefont
  {Martín-Martínez}}, \bibinfo {author} {\bibfnamefont {M.}~\bibnamefont
  {Montero}}, \ and\ \bibinfo {author} {\bibfnamefont {M.}~\bibnamefont {del
  Rey}},\ }\href {\doibase 10.1103/PhysRevD.87.064038} {\bibfield  {journal}
  {\bibinfo  {journal} {Physical Review D}\ }\textbf {\bibinfo {volume} {87}},\
  \bibinfo {pages} {064038} (\bibinfo {year} {2013}{\natexlab{b}})}\BibitemShut
  {NoStop}%
\bibitem [{\citenamefont {Benincasa}\ \emph {et~al.}(2014)\citenamefont
  {Benincasa}, \citenamefont {Borsten}, \citenamefont {Buck},\ and\
  \citenamefont {Dowker}}]{benincasa_quantum_2014}%
  \BibitemOpen
  \bibfield  {author} {\bibinfo {author} {\bibfnamefont {D.~M.~T.}\
  \bibnamefont {Benincasa}}, \bibinfo {author} {\bibfnamefont {L.}~\bibnamefont
  {Borsten}}, \bibinfo {author} {\bibfnamefont {M.}~\bibnamefont {Buck}}, \
  and\ \bibinfo {author} {\bibfnamefont {F.}~\bibnamefont {Dowker}},\ }\href
  {\doibase 10.1088/0264-9381/31/7/075007} {\bibfield  {journal} {\bibinfo
  {journal} {Classical and Quantum Gravity}\ }\textbf {\bibinfo {volume}
  {31}},\ \bibinfo {pages} {075007} (\bibinfo {year} {2014})}\BibitemShut
  {NoStop}%
\bibitem [{\citenamefont {Jonsson}\ \emph {et~al.}(2014)\citenamefont
  {Jonsson}, \citenamefont {Martín-Martínez},\ and\ \citenamefont
  {Kempf}}]{jonsson_quantum_2014}%
  \BibitemOpen
  \bibfield  {author} {\bibinfo {author} {\bibfnamefont {R.~H.}\ \bibnamefont
  {Jonsson}}, \bibinfo {author} {\bibfnamefont {E.}~\bibnamefont
  {Martín-Martínez}}, \ and\ \bibinfo {author} {\bibfnamefont
  {A.}~\bibnamefont {Kempf}},\ }\href {\doibase 10.1103/PhysRevA.89.022330}
  {\bibfield  {journal} {\bibinfo  {journal} {Physical Review A}\ }\textbf
  {\bibinfo {volume} {89}},\ \bibinfo {pages} {022330} (\bibinfo {year}
  {2014})}\BibitemShut {NoStop}%
\bibitem [{\citenamefont {Eberhard}\ and\ \citenamefont
  {Ross}(1989)}]{eberhard_quantum_1989}%
  \BibitemOpen
  \bibfield  {author} {\bibinfo {author} {\bibfnamefont {P.~H.}\ \bibnamefont
  {Eberhard}}\ and\ \bibinfo {author} {\bibfnamefont {R.~R.}\ \bibnamefont
  {Ross}},\ }\href {\doibase 10.1007/BF00696109} {\bibfield  {journal}
  {\bibinfo  {journal} {Foundations of Physics Letters}\ }\textbf {\bibinfo
  {volume} {2}},\ \bibinfo {pages} {127} (\bibinfo {year} {1989})}\BibitemShut
  {NoStop}%
\bibitem [{\citenamefont {Fulling}(1989)}]{fulling_aspects_1989}%
  \BibitemOpen
  \bibfield  {author} {\bibinfo {author} {\bibfnamefont {S.~A.}\ \bibnamefont
  {Fulling}},\ }\href@noop {} {\emph {\bibinfo {title} {Aspects of {Quantum}
  {Field} {Theory} in {Curved} {Spacetime}}}}\ (\bibinfo  {publisher}
  {Cambridge University Press},\ \bibinfo {year} {1989})\BibitemShut {NoStop}%
\bibitem [{\citenamefont
  {Martín-Martínez}(2015)}]{martin-martinez_causality_2015}%
  \BibitemOpen
  \bibfield  {author} {\bibinfo {author} {\bibfnamefont {E.}~\bibnamefont
  {Martín-Martínez}},\ }\href {\doibase 10.1103/PhysRevD.92.104019}
  {\bibfield  {journal} {\bibinfo  {journal} {Physical Review D}\ }\textbf
  {\bibinfo {volume} {92}},\ \bibinfo {pages} {104019} (\bibinfo {year}
  {2015})}\BibitemShut {NoStop}%
\bibitem [{\citenamefont {Birrell}(1982)}]{birrell_quantum_1982}%
  \BibitemOpen
  \bibfield  {author} {\bibinfo {author} {\bibfnamefont {N.~D.}\ \bibnamefont
  {Birrell}},\ }\href@noop {} {\emph {\bibinfo {title} {Quantum fields in
  curved space}}},\ Cambridge monographs on mathematical physics ; [7]\
  (\bibinfo  {publisher} {Cambridge University Press},\ \bibinfo {address}
  {Cambridge Cambridgeshire},\ \bibinfo {year} {1982})\BibitemShut {NoStop}%
\bibitem [{\citenamefont {Wald}(1994)}]{wald_quantum_1994}%
  \BibitemOpen
  \bibfield  {author} {\bibinfo {author} {\bibfnamefont {R.~M.}\ \bibnamefont
  {Wald}},\ }\href@noop {} {\emph {\bibinfo {title} {Quantum field theory in
  curved spacetime and black hole thermodynamics}}},\ Chicago lectures in
  physics\ (\bibinfo  {publisher} {University of Chicago Press},\ \bibinfo
  {address} {Chicago},\ \bibinfo {year} {1994})\BibitemShut {NoStop}%
\bibitem [{\citenamefont {Lin}\ and\ \citenamefont
  {Hu}(2006)}]{lin_accelerated_2006}%
  \BibitemOpen
  \bibfield  {author} {\bibinfo {author} {\bibfnamefont {S.-Y.}\ \bibnamefont
  {Lin}}\ and\ \bibinfo {author} {\bibfnamefont {B.~L.}\ \bibnamefont {Hu}},\
  }\href {\doibase 10.1103/PhysRevD.73.124018} {\bibfield  {journal} {\bibinfo
  {journal} {Physical Review D}\ }\textbf {\bibinfo {volume} {73}},\ \bibinfo
  {pages} {124018} (\bibinfo {year} {2006})}\BibitemShut {NoStop}%
\bibitem [{\citenamefont {Kim}(1999)}]{kim_quantum_1999}%
  \BibitemOpen
  \bibfield  {author} {\bibinfo {author} {\bibfnamefont {H.-C.}\ \bibnamefont
  {Kim}},\ }\href {\doibase 10.1103/PhysRevD.59.064024} {\bibfield  {journal}
  {\bibinfo  {journal} {Physical Review D}\ }\textbf {\bibinfo {volume} {59}},\
  \bibinfo {pages} {064024} (\bibinfo {year} {1999})}\BibitemShut {NoStop}%
\bibitem [{\citenamefont {Kim}\ and\ \citenamefont
  {Kim}(1997)}]{kim_radiation_1997}%
  \BibitemOpen
  \bibfield  {author} {\bibinfo {author} {\bibfnamefont {H.-C.}\ \bibnamefont
  {Kim}}\ and\ \bibinfo {author} {\bibfnamefont {J.~K.}\ \bibnamefont {Kim}},\
  }\href {\doibase 10.1103/PhysRevD.56.3537} {\bibfield  {journal} {\bibinfo
  {journal} {Physical Review D}\ }\textbf {\bibinfo {volume} {56}},\ \bibinfo
  {pages} {3537} (\bibinfo {year} {1997})}\BibitemShut {NoStop}%
\bibitem [{\citenamefont {Massar}\ and\ \citenamefont
  {Parentani}(1996)}]{massar_vacuum_1996}%
  \BibitemOpen
  \bibfield  {author} {\bibinfo {author} {\bibfnamefont {S.}~\bibnamefont
  {Massar}}\ and\ \bibinfo {author} {\bibfnamefont {R.}~\bibnamefont
  {Parentani}},\ }\href {\doibase 10.1103/PhysRevD.54.7426} {\bibfield
  {journal} {\bibinfo  {journal} {Physical Review D}\ }\textbf {\bibinfo
  {volume} {54}},\ \bibinfo {pages} {7426} (\bibinfo {year}
  {1996})}\BibitemShut {NoStop}%
\bibitem [{\citenamefont {Audretsch}\ and\ \citenamefont
  {Müller}(1994)}]{audretsch_radiation_1994}%
  \BibitemOpen
  \bibfield  {author} {\bibinfo {author} {\bibfnamefont {J.}~\bibnamefont
  {Audretsch}}\ and\ \bibinfo {author} {\bibfnamefont {R.}~\bibnamefont
  {Müller}},\ }\href {\doibase 10.1103/PhysRevD.49.6566} {\bibfield  {journal}
  {\bibinfo  {journal} {Physical Review D}\ }\textbf {\bibinfo {volume} {49}},\
  \bibinfo {pages} {6566} (\bibinfo {year} {1994})}\BibitemShut {NoStop}%
\bibitem [{\citenamefont {Louko}\ and\ \citenamefont
  {Satz}(2008)}]{louko_transition_2008}%
  \BibitemOpen
  \bibfield  {author} {\bibinfo {author} {\bibfnamefont {J.}~\bibnamefont
  {Louko}}\ and\ \bibinfo {author} {\bibfnamefont {A.}~\bibnamefont {Satz}},\
  }\href {\doibase 10.1088/0264-9381/25/5/055012} {\bibfield  {journal}
  {\bibinfo  {journal} {Classical and Quantum Gravity}\ }\textbf {\bibinfo
  {volume} {25}},\ \bibinfo {pages} {055012} (\bibinfo {year} {2008})},\
  \bibinfo {note} {arXiv: 0710.5671}\BibitemShut {NoStop}%
\bibitem [{\citenamefont {Satz}(2007)}]{satz_then_2007}%
  \BibitemOpen
  \bibfield  {author} {\bibinfo {author} {\bibfnamefont {A.}~\bibnamefont
  {Satz}},\ }\href {\doibase 10.1088/0264-9381/24/7/003} {\bibfield  {journal}
  {\bibinfo  {journal} {Classical and Quantum Gravity}\ }\textbf {\bibinfo
  {volume} {24}},\ \bibinfo {pages} {1719} (\bibinfo {year} {2007})},\ \bibinfo
  {note} {arXiv: gr-qc/0611067 bibtex: satz\_then\_2007}\BibitemShut {NoStop}%
\bibitem [{\citenamefont {Takagi}(1986)}]{takagi_vacuum_1986}%
  \BibitemOpen
  \bibfield  {author} {\bibinfo {author} {\bibfnamefont {S.}~\bibnamefont
  {Takagi}},\ }\href {\doibase 10.1143/PTPS.88.1} {\bibfield  {journal}
  {\bibinfo  {journal} {Progress of Theoretical Physics Supplement}\ }\textbf
  {\bibinfo {volume} {88}},\ \bibinfo {pages} {1} (\bibinfo {year}
  {1986})}\BibitemShut {NoStop}%
\bibitem [{\citenamefont {Juárez-Aubry}\ and\ \citenamefont
  {Louko}(2014)}]{juarez-aubry_onset_2014}%
  \BibitemOpen
  \bibfield  {author} {\bibinfo {author} {\bibfnamefont {B.~A.}\ \bibnamefont
  {Juárez-Aubry}}\ and\ \bibinfo {author} {\bibfnamefont {J.}~\bibnamefont
  {Louko}},\ }\href {\doibase 10.1088/0264-9381/31/24/245007} {\bibfield
  {journal} {\bibinfo  {journal} {Classical and Quantum Gravity}\ }\textbf
  {\bibinfo {volume} {31}},\ \bibinfo {pages} {245007} (\bibinfo {year}
  {2014})}\BibitemShut {NoStop}%
\bibitem [{\citenamefont {Martín-Martínez}\ and\ \citenamefont
  {Louko}(2014)}]{martin-martinez_particle_2014}%
  \BibitemOpen
  \bibfield  {author} {\bibinfo {author} {\bibfnamefont {E.}~\bibnamefont
  {Martín-Martínez}}\ and\ \bibinfo {author} {\bibfnamefont {J.}~\bibnamefont
  {Louko}},\ }\href {\doibase 10.1103/PhysRevD.90.024015} {\bibfield  {journal}
  {\bibinfo  {journal} {Physical Review D}\ }\textbf {\bibinfo {volume} {90}},\
  \bibinfo {pages} {024015} (\bibinfo {year} {2014})}\BibitemShut {NoStop}%
\bibitem [{\citenamefont {Brown}\ \emph {et~al.}(2013)\citenamefont {Brown},
  \citenamefont {Martín-Martínez}, \citenamefont {Menicucci},\ and\
  \citenamefont {Mann}}]{brown_detectors_2013}%
  \BibitemOpen
  \bibfield  {author} {\bibinfo {author} {\bibfnamefont {E.~G.}\ \bibnamefont
  {Brown}}, \bibinfo {author} {\bibfnamefont {E.}~\bibnamefont
  {Martín-Martínez}}, \bibinfo {author} {\bibfnamefont {N.~C.}\ \bibnamefont
  {Menicucci}}, \ and\ \bibinfo {author} {\bibfnamefont {R.~B.}\ \bibnamefont
  {Mann}},\ }\href {\doibase 10.1103/PhysRevD.87.084062} {\bibfield  {journal}
  {\bibinfo  {journal} {Physical Review D}\ }\textbf {\bibinfo {volume} {87}},\
  \bibinfo {pages} {084062} (\bibinfo {year} {2013})}\BibitemShut {NoStop}%
\bibitem [{\citenamefont {Bruschi}\ \emph {et~al.}(2013)\citenamefont
  {Bruschi}, \citenamefont {Lee},\ and\ \citenamefont
  {Fuentes}}]{bruschi_time_2013}%
  \BibitemOpen
  \bibfield  {author} {\bibinfo {author} {\bibfnamefont {D.~E.}\ \bibnamefont
  {Bruschi}}, \bibinfo {author} {\bibfnamefont {A.~R.}\ \bibnamefont {Lee}}, \
  and\ \bibinfo {author} {\bibfnamefont {I.}~\bibnamefont {Fuentes}},\ }\href
  {\doibase 10.1088/1751-8113/46/16/165303} {\bibfield  {journal} {\bibinfo
  {journal} {Journal of Physics A: Mathematical and Theoretical}\ }\textbf
  {\bibinfo {volume} {46}},\ \bibinfo {pages} {165303} (\bibinfo {year}
  {2013})}\BibitemShut {NoStop}%
\bibitem [{\citenamefont {Adesso}\ \emph {et~al.}(2014)\citenamefont {Adesso},
  \citenamefont {Ragy},\ and\ \citenamefont {Lee}}]{adesso_continuous_2014}%
  \BibitemOpen
  \bibfield  {author} {\bibinfo {author} {\bibfnamefont {G.}~\bibnamefont
  {Adesso}}, \bibinfo {author} {\bibfnamefont {S.}~\bibnamefont {Ragy}}, \ and\
  \bibinfo {author} {\bibfnamefont {A.~R.}\ \bibnamefont {Lee}},\ }\href
  {\doibase 10.1142/S1230161214400010} {\bibfield  {journal} {\bibinfo
  {journal} {arXiv:1401.4679 [cond-mat, physics:math-ph, physics:physics,
  physics:quant-ph]}\ } (\bibinfo {year} {2014}),\ 10.1142/S1230161214400010},\
  \bibinfo {note} {arXiv: 1401.4679}\BibitemShut {NoStop}%
\bibitem [{\citenamefont {Braunstein}\ and\ \citenamefont {van
  Loock}(2005)}]{braunstein_quantum_2005}%
  \BibitemOpen
  \bibfield  {author} {\bibinfo {author} {\bibfnamefont {S.~L.}\ \bibnamefont
  {Braunstein}}\ and\ \bibinfo {author} {\bibfnamefont {P.}~\bibnamefont {van
  Loock}},\ }\href {\doibase 10.1103/RevModPhys.77.513} {\bibfield  {journal}
  {\bibinfo  {journal} {Reviews of Modern Physics}\ }\textbf {\bibinfo {volume}
  {77}},\ \bibinfo {pages} {513} (\bibinfo {year} {2005})}\BibitemShut
  {NoStop}%
\bibitem [{\citenamefont {Weedbrook}\ \emph {et~al.}(2012)\citenamefont
  {Weedbrook}, \citenamefont {Pirandola}, \citenamefont {García-Patrón},
  \citenamefont {Cerf}, \citenamefont {Ralph}, \citenamefont {Shapiro},\ and\
  \citenamefont {Lloyd}}]{weedbrook_gaussian_2012}%
  \BibitemOpen
  \bibfield  {author} {\bibinfo {author} {\bibfnamefont {C.}~\bibnamefont
  {Weedbrook}}, \bibinfo {author} {\bibfnamefont {S.}~\bibnamefont
  {Pirandola}}, \bibinfo {author} {\bibfnamefont {R.}~\bibnamefont
  {García-Patrón}}, \bibinfo {author} {\bibfnamefont {N.~J.}\ \bibnamefont
  {Cerf}}, \bibinfo {author} {\bibfnamefont {T.~C.}\ \bibnamefont {Ralph}},
  \bibinfo {author} {\bibfnamefont {J.~H.}\ \bibnamefont {Shapiro}}, \ and\
  \bibinfo {author} {\bibfnamefont {S.}~\bibnamefont {Lloyd}},\ }\href
  {\doibase 10.1103/RevModPhys.84.621} {\bibfield  {journal} {\bibinfo
  {journal} {Reviews of Modern Physics}\ }\textbf {\bibinfo {volume} {84}},\
  \bibinfo {pages} {621} (\bibinfo {year} {2012})}\BibitemShut {NoStop}%
\bibitem [{\citenamefont {Jonsson}(2016)}]{jonsson_decoupling_2016}%
  \BibitemOpen
  \bibfield  {author} {\bibinfo {author} {\bibfnamefont {R.~H.}\ \bibnamefont
  {Jonsson}},\ }\emph {\bibinfo {title} {Decoupling of {{Information
  Propagation}} from {{Energy Propagation}}}},\ \href@noop {} {Ph.D. thesis},\
  \bibinfo  {school} {University of Waterloo} (\bibinfo {year}
  {2016})\BibitemShut {NoStop}%
\bibitem [{\citenamefont {Lloyd}\ \emph {et~al.}(2015)\citenamefont {Lloyd},
  \citenamefont {Chiloyan}, \citenamefont {Hu}, \citenamefont {Huberman},
  \citenamefont {Liu},\ and\ \citenamefont {Chen}}]{lloyd_no_2015}%
  \BibitemOpen
  \bibfield  {author} {\bibinfo {author} {\bibfnamefont {S.}~\bibnamefont
  {Lloyd}}, \bibinfo {author} {\bibfnamefont {V.}~\bibnamefont {Chiloyan}},
  \bibinfo {author} {\bibfnamefont {Y.}~\bibnamefont {Hu}}, \bibinfo {author}
  {\bibfnamefont {S.}~\bibnamefont {Huberman}}, \bibinfo {author}
  {\bibfnamefont {Z.-W.}\ \bibnamefont {Liu}}, \ and\ \bibinfo {author}
  {\bibfnamefont {G.}~\bibnamefont {Chen}},\ }\href
  {http://arxiv.org/abs/1510.05035} {\bibfield  {journal} {\bibinfo  {journal}
  {arXiv:1510.05035 [quant-ph]}\ } (\bibinfo {year} {2015})},\ \bibinfo {note}
  {arXiv: 1510.05035}\BibitemShut {NoStop}%
\end{thebibliography}%

\end{document}